\documentclass[twocolumn,
 amsmath,
 amssymb,
 aps,
 nofootinbib,
 preprintnumbers,
 floatfix
]{revtex4-1}
\usepackage[utf8]{inputenc}
\usepackage{simplewick}
\usepackage{footnote}
\usepackage{graphicx}
\usepackage{dcolumn}
\usepackage{bm}
\usepackage{stmaryrd}
\usepackage{amsmath}
\usepackage{amssymb}
\usepackage{bbm}
\usepackage{amsfonts}
\usepackage{xcolor}
\usepackage{footmisc}
\usepackage{comment}
\usepackage{ulem}
\usepackage{cancel}
\usepackage{nicefrac}
\usepackage[linktocpage=true,bookmarksnumbered=true,colorlinks=true,plainpages,linkcolor=blue,citecolor=red]{hyperref}

\begin{document}
\preprint{ADP-23-20/T1229, CPPC-2023-01}
\title{Revisiting Cosmological Constraints on Supersymmetric SuperWIMPs}

\author{Meera~Deshpande$^{a}$}
\author{Jan~Hamann$^{b}$}
\author{Dipan~Sengupta$^{a}$}
\author{Martin~White$^{a}$}
\author{Anthony~G.~Williams$^{a}$}
\author{Yvonne~Y.~Y.~Wong$^{b}$}

 \affiliation{$^a$ARC Centre of Excellence for Dark Matter Particle Physics, Department of Physics, The University of Adelaide, Adelaide SA 5005, Australia}
 \affiliation{$^{b}$Sydney Consortium for Particle Physics and Cosmology, School of Physics, The University of New South Wales, Sydney NSW 2052, Australia}
 \begin{abstract}
    SuperWIMPs are extremely weakly interacting massive particles that inherit their relic abundance from late decays of frozen-out parent particles. Within supersymmetric models, gravitinos and axinos represent two of the most well motivated superWIMPs. In this paper we revisit constraints on these scenarios from a variety of cosmological observations that probe their production mechanisms as well as the superWIMP kinematic properties in the early Universe.  We consider in particular observables of Big Bang Nucleosynthesis and the Cosmic Microwave Background (spectral distortion and anisotropies), which limit the fractional energy injection from the late decays, as well as warm and mixed dark matter constraints derived from the Lyman-$\alpha$ forest and other small-scale structure observables.  We discuss complementary constraints from collider experiments,
    and argue that cosmological considerations rule out a significant part of the gravitino and the axino superWIMP parameter space.
    \end{abstract}

\maketitle

\section{Introduction}

The pursuit of signatures of beyond-the-Standard-Model (BSM) physics and an explanation for the dark matter of the Universe has been the holy grail for particle physicists for over three decades. To this end, the Large Hadron Collider (LHC) has probed large swathes of parameter space in a variety of well motivated BSM models. These include 
Supersymmetry (SUSY), the leading BSM theory that not only solves the hierarchy problem but also provides a slew of particle candidates for the dark matter. Within the context of specific SUSY breaking scenarios 
as well as simplified models, null measurements at the LHC have translated into constraints on a significant chunk of SUSY particles in the GeV-to-TeV mass range~\cite{ATLAS:2023afl,ATLAS:2023lfr,CMS-PAS-SUS-19-001}. 

However, SUSY/BSM searches at the LHC rely primarily on prompt decays or, at best, decays with proper lengths of $\mathcal{O}(10-100)$~m in the so-called long-lived particle (LLP) 
searches~\cite{ATLAS:2023meo,ATLAS:2023oti}. These searches are further subject to the constraints that the produced particles are within kinematic reach of the LHC, i.e., their masses are at most a few TeV, and that they are produced with a cross-section sufficient to generate a detectable signal over the enormous Standard-Model background.   Extremely weakly-interacting particles and/or those with very long lifetimes---many of which also reside in well-motivated SUSY/BSM model and parameter spaces---are thus inherently out of the LHC's reach, even if their masses lie within the conventional GeV-to-TeV collider window.

Interestingly, when the proper decay lengths/lifetimes of these particles exceed $\mathcal{O}(10)$~m, a second, albeit less conventional, window to explore their properties opens up.  Disregarding concerns of naturalness, scenarios of extremely long particle lifetimes could be easily realised in a wide range of BSM theories by particle masses and couplings spanning orders of magnitude [e.g., $m \sim \rm \mathcal{O}(1)~MeV - \mathcal{O}(100)~TeV$].
In the context of SUSY, these scenarios fall under the superWIMP class of models~\cite{Feng:2003uy,Feng:2004mt}, wherein quasi-stable particles can be efficiently produced in the early Universe and decay at a very late time, i.e., $t \gg {\cal O}(1)$~s post-Big Bang, during the standard cosmological history. Regardless of whether these quasi-stable particles can account for all of the observed dark matter abundance of the Universe at early times, late-decaying particles leave potentially observable signatures in the cosmic microwave background radiation (CMB), as well as the light element abundances from Big Bang Nucleosynthesis (BBN) and the large-scale matter distribution, particularly the Lyman-$\alpha$ (Ly$\alpha$) forest. Measurements of these observables can in turn be used to probe and constrain regions of SUSY/BSM parameter space inaccessible to collider searches. 

As a concrete example, consider the following: in $R$-parity-conserving SUSY, the conventional lightest SUSY particle (LSP) dark matter candidate is the lightest neutralino, superpartner of the electroweak gauge particles.  With masses  $m_{\rm LSP} \sim {\cal O}(0.1-1)$~TeV and weak-like interactions with the Standard Model (SM), the neutralino easily satisfies the observed relic abundance of the Universe and has a range of signatures at collider physics experiments, as well as at direct and indirect dark matter searches~\cite{Abdallah:2015ter,Bramante:2015una}. However, in models of Supergravity or in SUSY models extended to include the axion, the lightest neutralino may not be the LSP but can decay to lighter particles of these theories, such as the gravitino~$\tilde{G}$ or the axino~$\tilde{a}$.  That is, the neutralino is now the next-to-lightest supersymmetric particle (NLSP), while the gravitino or the axino serves as the LSP~\cite{Feng:2003uy,Feng:2004mt,Covi:2001nw,Covi:2004rb,Covi:2009bk}. 

In the latter scenarios, the decay widths of the NLSP neutralino to $\tilde{G}$ and $\tilde{a}$ are generally suppressed, either by the Planck mass $m_{\rm Pl}$ or by the axion decay constant $f_a$, such that the lifetime of the NLSP can be much longer than its freeze-out time scale.  In this case, the decay of the NLSP can also generate an axino or gravitino population.  Provided that the reheating temperature is low enough to avoid significant production of $\tilde{G}$ or $\tilde{a}$ from thermal scattering in the very early~Universe~\cite{Rychkov:2007uq}, it is the late-time NLSP-to-LSP decay process that dominates the final $\tilde{G}$ or $\tilde{a}$ abundance.  

Then, the relic LSP production can be thought of as  a two-stage process. First, a neutralino NLSP population is produced by interactions with the SM.%
\footnote{The mechanics of this production depends on the nature of the neutralino. For a light neutralino $\chi_{1}^{0} \lesssim 100$~GeV, one requires a bino-like neutralino to have enough annihilations to avoid overclosing the Universe~\cite{Barman:2017swy,Roszkowski:2017nbc}.}
Such a neutralino population with the right observed relic density can build up either via the usual thermal freeze-out mechanism through annihilations with SM particles, or via freeze-in if extremely-weakly coupled to the parent SUSY and other SM particles. Irrespective of the details of this first step, at very late times the neutralino NLSP decays into the gravitino or axino LSP, constituting the second step of the LSP production process.  The gravitino or axino LSP thus generated---dubbed as SUSY superWIMPs in the literature---can provide part or all of the observed dark matter in the Universe~\cite{Feng:2003xh,Feng:2003uy,Feng:2010gw,Feng:2010ij}.

Because superWIMPs are extremely weakly coupled---the interactions of the gravitino and axino are  $m_{\rm Pl}$- or $f_a$-suppressed---prompt searches at colliders are insensitive to a large part of their parameter space.  Only a small sliver can potentially be probed~\cite{Arbey:2015vlo,Kim:2019vcp,Heisig:2013sva,Cahill-Rowley:2012ydr,Maltoni:2015twa,Brandenburg:2005he,Feng:2010ij,Covi:2004rb,Choi:2011yf,Choi:2013lwa,Covi:2009bk,Covi:2009pq,Co:2016fln,Choi:2013lwa,Cheung:2011mg},
via searches for long-lived particles by ATLAS, CMS, or future experiments such as FASER~\cite{Feng:2022inv} and Mathusla~\cite{Curtin:2018mvb}.%
\footnote{This statement assumes the neutralino $\chi_{1}^{0}$ has not been ruled out by prompt jets/leptons + missing energy searches at the LHC. Within the general 19-parameter phenomenological MSSM (PMSSM) scenario~\cite{Chung:2003fi}, large regions of the parameter space are unconstrained~\cite{GAMBIT:2017zdo}. This is discussed in more detail in Sec.~\ref{sec:SUSYsuperWIMPs}.}
In a similar vein, bounds on the superWIMP parameter space from direct and indirect dark matter searches are practically non-existent.  Our best prospects for probing and constraining superWIMP scenarios lie in cosmological observations.

A number of early studies have considered how gravitino superWIMPs can be probed cosmologically~\cite{Pradler:2006hh,Steffen:2006hw,Feng:2004mt,Feng:2003uy,Steffen:2008qp,Cahill-Rowley:2012ydr}, based primarily on the premise that electromagnetic and/or hadronic energy released from the NLSP-to-LSP decay has consequences for the light element yields from BBN and the CMB black-body energy spectrum.  Using measurements of the Deuterium and Helium-4 abundances, as well as the COBE-FIRAS constraint on $\mu$-type spectral distortions, stringent constraints on the gravitino superWIMP parameter space can be set for NLSP lifetimes in the range $t_{\rm NLSP} \sim 10^4 - 10^8$~s.
Similar considerations have also been applied to the axino superWIMP scenario, wherein a frozen-out neutralino or stau decays to an axino accompanied by an SM particle~\cite{Covi:2001nw,Choi:2013lwa,Choi:2011yf,Covi:2009pq,Steffen:2008qp,Brandenburg:2005he,Covi:2004rb,Freitas:2009jb}.
Independently of whether a particular axion-axino superfield realisation solves the strong CP problem, if the axino is the LSP, cosmological data can be expected to constrain a large part of the parameter space.

In this work, we extend these early analyses to include constraints from the CMB temperature and polarisation anisotropies from the \textit{Planck} CMB mission~\cite{Planck:2018vyg}.  Just as they impact on the light elements and the CMB energy spectrum, electromagnetic energy injections from NLSP decay can likewise have drastic consequences for the reionisation history of the Universe.  Energy injection over NLSP lifetimes of $t_{\rm NLSP} \sim 10^{10} - 10^{24}$~s, in particular, can significantly modify the evolution of the free-electron fraction in the cosmic plasma, altering the CMB anisotropies in ways that are strongly disfavoured by current anisotropy measurements~\cite{Planck:2018vyg,Lucca:2019rxf}.
As we shall demonstrate, this in turn allows us to place stringent constraints on large swathes of superWIMP parameter space previously considered viable,
providing a powerful complement to conventional SUSY dark matter searches at colliders as well as at direct and indirect dark matter detection experiments.%
\footnote{We note that, in the recent years, several works have also invoked late-decaying dark matter to explain cosmological anomalies such as the Hubble and the $\sigma_{8}$ tensions~\cite{Gu:2020ozv,Choi:2021uhy,Hamaguchi:2017ihw}.  Most of these scenarios do not generate significant electromagnetic emissions and are thus not subject energy injection constraints.  The dominant constraints applicable to these scenarios arise from free-streaming effects and/or radiation excess (i.e., a non-standard $N_{\rm eff}$ relative to the Standard-Model prediction of $N_{\rm eff}^{\rm SM}=3.0440 \pm 0.0002$~\cite{Bennett:2020zkv,Bennett:2019ewm,Froustey:2020mcq}.}

 The paper is organised as follows.  We begin in Sec.~\ref{sec:SUSYsuperWIMPs} with a summary of some well-motivated SUSY superWIMP scenarios amenable to the observational and experimental constraints of this work. In Sec.~\ref{sec:constraints} we describe how these constraints can be applied to derive limits on the superWIMP parameter space, starting with cosmological observations and concluding with collider searches. Sec.~\ref{sec:results} summarises the limits thus derived on the gravitino and axino superWIMP parameter space.  We conclude in Sec.~\ref{sec:conclusions}. Appendix~\ref{sec:appendixA} outlines the derivation of the LSP momentum distribution expected from NLSP-to-LSP decay.

\section{SUSY superWIMPs}
\label{sec:SUSYsuperWIMPs}

The general mechanism of superWIMP production in the early Universe is straightforward. Heavier SUSY particles undergo cascade decays to lighter SUSY particles and eventually to the NLSP. The NLSP then freezes out,   typically at $x_{\rm f} \equiv m_{\rm NLSP}/T\sim 25$--30,
with a yield $Y_{\rm NLSP} \equiv  n_{\rm NLSP}/s$, where $n_{\rm NLSP}$ is the NLSP number density and $s$ is the entropy. Long after freeze-out, the NLSP decays to the LSP. Assuming the decay is complete,  $Y_{\rm LSP} \simeq Y_{\rm NLSP}$, and the final LSP abundance is simply given by $\Omega_{\rm LSP} h^{2} \simeq (m_{\rm LSP}/m_{\rm NLSP})\, \Omega_{\rm NLSP} h^{2}$, where $h$ is the reduced Hubble parameter. In most superWIMP scenarios, $m_{\rm NLSP}\simeq m_{\rm LSP}$, such that LSP inherits the same abundance as the NLSP. A large mass difference is however not precluded and can be a means to relax constraints on the LSP parameter space from relic density considerations. 

For SUSY superWIMPs, if there exists a thermal production mechanism in the early Universe generating an abundance proportional to the reheating temperature, then the total superWIMP abundance today is simply the sum of the thermal population and the population arising from NLSP decay (``non-thermal''), i.e.,
\begin{equation}
    \Omega_\mathrm{LSP}h^{2} = \Omega_\mathrm{LSP}^\mathrm{thermal}h^{2} + \Omega_\mathrm{LSP}^\mathrm{non-thermal}h^{2}.
\end{equation}
In general, however, to generate by thermal means a GeV-to-TeV-mass SUSY superWIMP population to match the observed dark matter abundance requires a reheating temperature in excess of $T_{\rm rh} \sim 10^{10}$~GeV~\cite{Rychkov:2007uq}.  Thus, if the reheating temperature turns out to be low, production of superWIMP relics must rely entirely on NLSP-to-LSP decay.

At this stage it is important to emphasise that, within the general Minimal Supersymmetric Standard Model (MSSM), the mechanism of thermal neutralino freeze-out that generates the right relic abundance is quite restricted given collider and electroweak precision observables, as well as constraints on the Higgs and $Z$-boson invisible widths \cite{Belanger:2013pna,Roszkowski:2017nbc,Barman:2017swy,Barman:2022jdg,AlbornozVasquez:2010nkq}.
The mechanism of thermal freeze-out for a relic neutralino depends on the nature of the gauge composition of the neutralino; 
for a comprehensive recent summary, see, e.g.,~\cite{Roszkowski:2017nbc}.
If the neutralino is light, i.e., $\rm m_{\chi_{1}^{0}}\lesssim 100 ~GeV$, limits on the charged components of the neutralino sector demand that 
the light neutral component $\chi_{1}^{0}$  be predominantly bino.  Then, 
imposing the \textit{Planck}-inferred dark matter density, $\Omega_{\rm DM} h^{2}\lesssim 0.12$ \cite{Planck:2018vyg}, on the neutral relic density leads immediately to a lower limit of $m_{\chi_{1}^{0}}\gtrsim 34$ GeV on the neutralino mass~\cite{Barman:2017swy}.
Thus, on the light neutralino side, assuming a thermal freeze-out mechanism the two places with maximally efficient enhancements in the annihilation cross-section so as not to overclose the Universe are at the $Z$-funnel and the Higgs funnel regions~\cite{Barman:2017swy}.%
\footnote{This rather stringent condition on the neutralino mass can be relaxed within the next-to-Minimal Supersymmetric Standard Model (NMSSM), where the presence of additional scalars ensure an efficient annihilation~\cite{Barman:2020vzm}. Alternatively,
a non-thermal neutralino will also ensure that these limits are significantly weakened~\cite{Barman:2017swy}.}
Direct detection constraints however rule out a significant part of the neutralino parameter space in the 10~GeV-to-1~TeV mass range~\cite{Barman:2017swy,Mohan:2019zrk,GAMBIT:2017zdo,Bramante:2015una}, with limits depending on the specifics of the model parameters.  In general spin-independent limits from the Xenon-nT/LZ direct detection  experiments~\cite{Barman:2022jdg} are quite constraining in the light dark matter scenario ($\rm m_{\chi_{1}^{0}}\lesssim 200~GeV$),  leaving viable the $Z/H$ funnel regions.%
\footnote{The limits are sensitive to the Higgsino mass parameter $\mu$. While the $\mu > 0$ parameter space is severely restricted, constraints on $\mu< 0$ are not as restrictive \cite{Barman:2017swy,Barman:2022jdg} }

At higher masses, depending on the gauge content of the neutralino and the SUSY mass spectrum, a variety of new 
annihilation mechanisms can 
open up. Given the strong limits from collider searches, the most promising scenarios proceed through co-annihilations with sleptons and squarks. For the latter, co-annihilations aided by Sommerfeld enhancements can lead to the correct relic density~\cite{Bramante:2015una,Becker:2022iso}. 
To briefly conclude this discussion we also add that if the neutralino has a sizable Higgsino component, 
a TeV scale Higgsino can generate the relic abundance of the Universe through  co-annihilation with nearly mass-degenerate charginos~\cite{Mizuta:1992qp}; this scenario requires the so-called ``well-tempered'' neutralino, a right admixture of bino and Higgsino for efficient annihilation~\cite{Mizuta:1992qp}.
Since the charginos  are TeV scale electroweak gauginos,  collider limits can be evaded if the rest of the SUSY spectrum is decoupled as in the Split SUSY cases~\cite{Giudice:2004tc}. These considerations are generally encoded within the idea of the so-called relic neutralino surface~\cite{Bramante:2015una}. General phenomenological and simplified MSSM model  studies for the electroweakino sector using LHC data have shown that large swathes of parameter space are allowed within the gaugino sector, implying that there is no generic model-independent lower bound on the light neutralino~\cite{GAMBIT:2017zdo}.  The situation is relaxed further in non-minimal models like the next-to-Minimal Supersymmetric Standard Model (NMSSM) or non-universal Gaugino Models (NUGM).
We also emphasise that, in models with over-abundant
dark matter (e.g., models involving a light bino-like neutralino), the superWIMP mechanism is a way to dilute the final relic abundance.

In what follows, we briefly describe two well-motivated SUSY superWIMPs, the gravitino~$\tilde{G}$ and the axino~$\tilde{a}$.
As we shall see in Sec.~\ref{sec:results}, irrespective of the freeze-out/freeze-in mechanism that produces the NLSP neutralino,  energy injection constraints of PMSSM and BBN, coupled with free-streaming bounds from the Ly$\alpha$ data will constrain the bulk of their parameter spaces. 
 

\subsection{Gravitino superWIMPs} 

Gravitinos $\tilde{G}$ are spin-$\nicefrac{3}{2}$ superpartners of gravitons.  Depending on the SUSY breaking mechanism, the gravitino mass---given approximately by $m_{\tilde{G}} \simeq \langle F \rangle/m_{\rm pl}$, where $\langle F \rangle$ is   the SUSY breaking scale---can range from keV to TeV and is thus essentially a free parameter in this study.  Because interactions of the gravitino are $m_{\rm Pl}$-suppressed, we do not expect them to be efficiently produced via scattering in the early Universe  unless the reheating temperature is high~\cite{Rychkov:2007uq}. 

Production from NLSP decay can proceed via the decay of the lightest neutralino $\chi_{1}^{0}$.
 Stringent BBN constraints on hadronic energy injection from the decays $\chi_{1}^{0}\to \tilde{G} h/Z$ essentially rules out a predominantly wino- or Higgsino-like neutralino~\cite{Feng:2003uy,Feng:2003xh}. Then, what remains is a bino-like neutralino, which decays into a gravitino predominantly via the two-body decay $\chi_{1}^{0}\to \tilde{G} \gamma$, whose width is given by~\cite{Feng:2003uy}
 \begin{equation}
\label{eq:GravitinoDecayWidth}
\Gamma(\chi_{1}^{0}\to \tilde{G} \gamma ) = \frac{m_{\chi_{1}^{0}}^{5} \cos^{2}\theta_\mathrm{W}}{48 { \pi} m_{\rm Pl}^{2}m_{\tilde{G}}^{2}}\left(1 - \frac{m_{\tilde{G}}^{2}}{m_{\chi_{1}^{0}}^{2}} \right)^{3}\!\! \left(1 + 3\frac{m_{\tilde{G}}^{2}}{m_{\chi_{1}^{0}}^{2}} \right),
\end{equation}
where $m_{Pl}$ is the reduced Planck mass, and $\theta_\mathrm{W}$ is the weak mixing angle.  

Assuming decay at rest and that the energy carried by the photon, $E_{\gamma}= (m_{\chi_{1}^{0}}^{2} - m_{\tilde{G}}^{2})/(2m_{\chi_{1}^{0}})$, is injected entirely into the cosmic plasma, it is convenient to recast the width~\eqref{eq:GravitinoDecayWidth} as 
\begin{equation}
\begin{aligned}
\label{eq:GravitinoDecayWidth2}
\Gamma(\chi_{1}^{0}\to \tilde{G} \gamma ) 
& = \frac{m_{\chi_{1}^{0}}^{3} \cos^{2}\theta_\mathrm{W}}{3  \pi m_{\rm Pl}^{2}}  \epsilon_{\rm em}^3 \frac{2-3 \epsilon_{\rm em}}{1- 2 \epsilon_{\rm em}} \\
&\simeq 2.2 \times 10^{-14}~{\rm s}^{-1} \,  \epsilon_{\rm em}^3 \frac{2-3 \epsilon_{\rm em}}{1- 2 \epsilon_{\rm em}} \left(\frac{m_{\chi_1^0}}{\rm GeV} \right)^3,
\end{aligned}
\end{equation}
with 
\begin{equation}
\label{eq:epsilonSM}
\epsilon_{\rm em} \equiv \frac{E_{\gamma}}{m_{\chi_{1}^{0}}}=\frac{ m_{\chi_{1}^{0}}^{2} - m_{\tilde{G}}^{2}}{2m^2_{\chi_{1}^{0}}}
\end{equation}
denoting the fraction of the neutralino mass released as electromagnetic energy. 
Where kinematically allowed, the additional decay channels $\chi_{1}^0 \to \tilde{G}Z/h$ are also available. But, as said above, these channels are suppressed for a bino-like $\chi^{0}_{1}$. Note that maximal energy injection is represented by $\epsilon_{\rm em}\to 0.5$, which occurs as $m_{\tilde{G}}\to 0$.%
\footnote{The limit $m_{\tilde{G}} = 0$ is ill-defined within theories of SUSY breaking mechanisms. Since $m_{\tilde{G}}$ is related to the SUSY breaking scale $\langle F\rangle$, the $\langle F\rangle \to 0$ limit simply means a decoupled massless gravitino. Swampland conjectures relate it to the massless limit of an infinite tower of states and the breakdown of the effective field theory~\cite{Cribiori:2021gbf}.}


\subsection{Axino superWIMPs}

Axinos $\tilde{a}$ are the supersymmetric partners of the axion---the dynamic field expected to solve the strong CP problem---and appear in the axion supermultiplet after the Peccei-Quinn (PQ) symmetry breaking in the form $A = (s +i a)/\sqrt{2} + \sqrt{2}\theta a + \theta^{2}F$, where $a$ is the axion, $s$ the saxion,%
\footnote{Although we will ignore the saxion for this work, saxions can also form superWIMPs and be subject to cosmological energy injection constraints.}
$F$ the auxiliary superfield, and $\theta$ is the Grassmanian coordinate.
The  axion couples derivatively to quarks and to the gauge bosons with interactions suppressed by the PQ breaking scale $f_{a}$; the accompanying SUSY interactions can be found by simply supersymmetrising the effective SM-axion interactions, i.e., the axion supermultiplet $A$ couples to the vector supermultiplet $V_{a}$. 
The axion supermultiplet acquires a mass after SUSY is broken. While the saxion mass is roughly set by the the soft SUSY breaking scale, the axino mass depends on the superpotential.
For the purposes of this work, we will take the axino mass to be a free parameter, and note that its mass can range from eV to TeV scales. 

Like gravitinos, axinos can be produced in the early Universe in abundance via thermal scattering if the reheating temperature is large~\cite{Strumia:2010aa}.  However, if the axino is the LSP, production from the decay of a NLSP neutralino population is also possible.  Assuming a (pure) bino decay, the decay width is given by \cite{Covi:2004rb,Covi:2009bk}
\begin{equation}
 \begin{aligned}
 \label{eq:decaywidthaxino}
  \Gamma \left(\chi_1^0 \to \tilde{a} \gamma\right)
& =   {\left(\frac{\alpha^{2}}{4\pi}\right)}  C_{aYY}^2 \frac{m_{\chi_{1}^{0}}^{3}}{4 \pi^{2}  {f'_{a}}^{2} \cos^{2}\theta_\mathrm{W}} \epsilon_{\rm em}^{3}\\
& \simeq  2.1 \times 10^{-15}~{\rm s}^{-1} \\
& \qquad \times C_{aYY}^2  \epsilon_{\rm em}^{3} \left(\frac{m_{\chi_1^0}}{\rm GeV} \right)^3 \left(\frac{{\rm GeV}}{f'_a/10^{16}} \right)^2.
\end{aligned}
\end{equation}
Here, $f'_{a} \equiv f_{a}/N$, where the factor $N=1$ and $N=6$ applies to the KSVZ and DFSZ axion, respectively; the coefficient $C_{aYY}$ is a model-dependent $\mathcal{O}(1)$ number~\cite{Covi:2004rb}, which we set to unity in this analysis without loss of generality; and $\epsilon_{\rm em}$ is given by Eq.~(\ref{eq:epsilonSM}), but with the replacement $m_{\tilde{G}} \to m_{\tilde{a}}$.
Precision cosmology currently limits the PQ breaking scale to $f_a \gtrsim 10^8$~GeV (via the axion hot dark matter fraction) for all axion models, while the DFSZ axion is further subject to red-giant constraints on the axion-electron coupling, such that $f_a \gtrsim 10^{10}$~GeV~\cite{ParticleDataGroup:2022pth}.  Note also that for $f_{a}\gtrsim 10^{12}$, the axion can contribute significantly to the observed dark matter abundance of the Universe.%
\footnote{If the  misalignment angle is ${\cal O}(1)$, such as in the post-inflationary scenario, then to explain the observed dark matter abundance of the Universe fixes $f_{a}\sim 10^{12}$~GeV.  However, in the ``anthropic'' or pre-inflationary scenario, the misalignment angle is random; in this case there is no upper limit on $f_a$.}

As in the case of the $\chi_1^0$ decay to gravitino, where kinematically viable, the decay $\chi_{1}^{0}\to \tilde{a} Z$ is also allowed albeit suppressed relative to $\chi_{1}^{0} \to \tilde{a} \gamma$ in both the decay width and the accompanying electromageatic energy release.
The possibility also exists that the gravitino (axino) is the NLSP and decays into the axino (gravitino) LSP accompanied by the release of an axion: this process has in fact been claimed to solve the Hubble tension through the injection of dark radiation~\cite{Hamaguchi:2017ihw}.


\section{Cosmological and collider probes of superWIMPs}
\label{sec:constraints}

From the cosmological perspective, the two defining features of superWIMPs are (i) the NLSP decays to the LSP on cosmological time scales, and (ii) the decay is accompanied by the release of electromagnetic radiation.  Irrespective of whether the NLSP or LSP accounts for the entirety of the observed dark matter abundance, these features can manifest themselves in precision cosmological observables either via the electromagnetic radiation or in the kinematic properties of the LSP itself.  We elaborate on the relevant cosmological observables and how they can be used to constrain superWIMPs in the following  subsections.  For completeness, we also discuss collider probes of superWIMPs in Sec.~\ref{sec:collider}.

\begin{figure}[t]
\includegraphics[width=.48\textwidth]{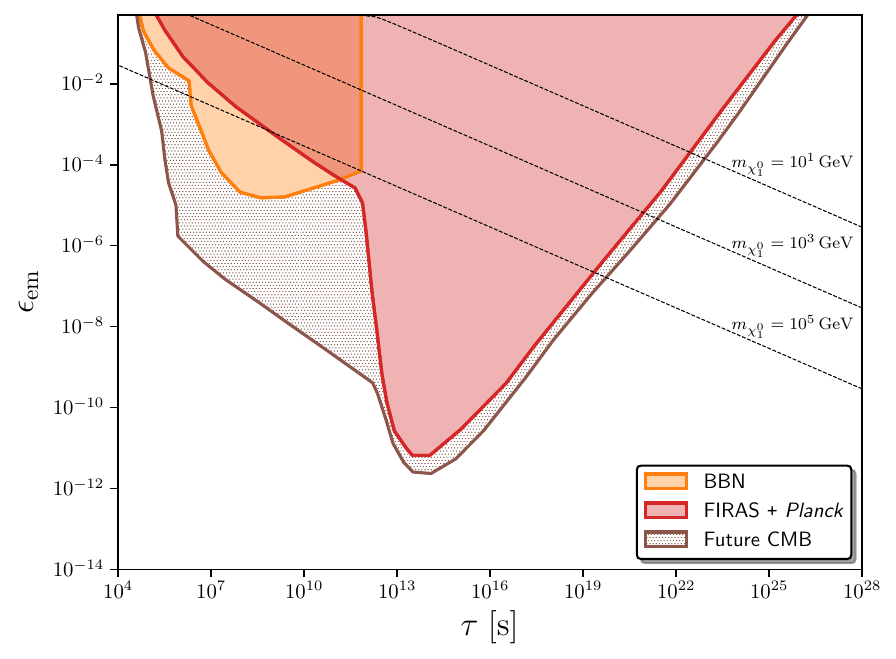}
    \caption{
    \label{Fig:gravEtau} 
    Current and projected constraints on the fractional energy injection~$\epsilon_{\rm em}$ as a function of decay lifetime~$\tau$ from the BBN light element abundances (H$^2$ and He$^3$), current CMB measurements (FIRAS\texttt{+}{\it Planck}), and future CMB probes (LiteBIRD\texttt{+}CMB-S4\texttt{+}PRISM).  These constraints have been extracted from Fig.~8 of Ref.~\cite{Lucca:2019rxf}, where the exclusion limits correspond to the $\Delta \chi=4$ isocontours on a variable fraction $f_{\rm frac}$ of dark matter decaying via ${\rm DM} \to \gamma \gamma$ for which the fractional injection is always fixed at $\epsilon_{\rm em}=1$; we have reintepreted these isocontours to constraints on a variable $\epsilon_{\rm em}$ for a fixed $f_{\rm frac}=1$. 
    The mapping is to a good approximation one-to-one, provided that the initial energy injection exceeds $\sim 100$~MeV in the case of BBN and $\sim 1$~MeV in the case of CMB.
    To illustrate the power of these constraints, the black dashed lines indicate predictions for the gravitino superWIMP scenario, based on Eq.~(\ref{eq:GravitinoDecayWidth2}), for several neutralino mass values.}
\end{figure}

\subsection{Light element abundances}
\label{sec:BBN}

The main effect of electromagnetic energy injection into the plasma on the abundances of the light elements (Deuterium $^2$H, Helium-3 $^3$He, and Helium-4 $^4$He) from BBN is photo-dissociation, provided that the injected energy exceeds the reaction threshold (typically ${\cal O}(2-30)$~MeV; see, e.g., Table~1 of Ref.~\cite{Acharya:2019uba}).  However, the energy fraction available for photo-dissociation is strongly suppressed if the injection occurs when the plasma temperature is above $T \sim 10$~keV (or, equivalently, $t \sim 10^4$~s).  This is because, at these plasma temperatures, energy injections that exceed even the lowest photo-dissociation thresholds will also exceed the threshold for pair production ($ \gamma \gamma \to e^+ e^-$), \mbox{$E_c \simeq m_e^2/(22 T)$}~\cite{Kawasaki:1994sc,Protheroe:1994dt}.  Given that CMB photons outnumber baryons by a factor of $10^9$, pair production must dominate over all other electromagnetic processes whenever kinematically allowed, and quickly degrade the injected energy to below the photo-dissociation threshold.  For this reason, as a tool for constraining electromagnetic energy injection from particle decay, the light element abundances are sensitive only to decay lifetimes longer than $\tau \sim 10^4$~s.

Figure~\ref{Fig:gravEtau} shows the region on the $(\tau,\epsilon_{\rm em})$-plane excluded by the $^2$H and $^3$He abundances (shaded orange).  These constraints have been extracted from Fig.~8 of Ref.~\cite{Lucca:2019rxf}, itself based on Ref.~\cite{Chluba:2013pya}'s adaption of Ref.~\cite{Kawasaki:2004qu}'s results, for initial energy injections larger than $\sim 100$~MeV.  Note that
Ref.~\cite{Lucca:2019rxf} had presented their results in the form of constraints on a variable $f_{\rm frac}$ as a function 
of the decay lifetime $\tau$, where $f_{\rm frac}$ is the fraction of dark matter that decays 
via ${\rm DM} \to \gamma \gamma$, i.e., the fractional energy injection is fixed at 
$\epsilon_{\rm em}=1$.  In Fig.~\ref{Fig:gravEtau} we have reinterpreted these constraints to be upper limits on the fractional energy injection $\epsilon_{\rm em}$ at a fixed $f_{\rm frac}=1$.  The reintepretation should to a good approximation be one-to-one, provided that, for the light element abundances, the initial energy injection always respects $\gtrsim 100$~MeV.  For sub-100 MeV injections, bounds on $f_{\rm frac}$ or $\epsilon_{\rm em}$ are strongly dependent on the initial energy injection~\cite{Poulin:2015opa,Poulin:2016anj,Acharya:2019uba}, making this mapping less straightforward.  The hard cut-off of the BBN limits at $\tau \simeq 10^{12}$~s in the figure is simply due to the lack of calculations in this parameter region in the literature; we are otherwise not aware of any fundamental reason why BBN constraints could not be extended to longer decay lifetimes.

To illustrate the power of energy injection constraints, we have also plotted in Fig.~\ref{Fig:gravEtau} predictions for the gravitino superWIMP scenario, based on Eq.~(\ref{eq:GravitinoDecayWidth2}), for a number of $\chi_{1}^{0}$ masses from $m_{\chi_1^0}=1$~GeV and up to the unitarity limit of $m_{\chi_1^0}=100$~TeV.%
\footnote{The unitarity limit quoted here comes from Ref.~\cite{Griest:1989wd}, which argues as follows: in order for the neutralino to not exceed the upper limit on the relic abundance, the velocity-averaged cross-section must satisfy $\langle\sigma v\rangle\gtrsim 3\times 10^{-26} \,\rm cm^{3}\, {\rm s}^{-1}$. For an $s$-wave annihilation with perturbative couplings, this imposes an upper limit on the neutralino mass of $m_{\chi_1^0}\lesssim 30 -100$~TeV, depending on the nature of the interaction.  We take $m_{\chi_1^0}\lesssim 100$~TeV as a conservative limit.\label{footnote:unitary}}
We emphasise that for a given mass hierarchy between the neutralino and the gravitino, the amount of energy released and the lifetime are fixed by Eq.~(\ref{eq:GravitinoDecayWidth2}). 
 That is, there are no further variables that can affect the gravitino superWIMP predictions in Fig.~\ref{Fig:gravEtau}. As mentioned earlier, the maximal fractional energy released can never be exactly equal to 0.5, which would correspond to a massless gravitino.  


\subsection{CMB spectral distortions}
\label{sec:spectdist}
Electromagnetic energy injection in the early Universe must also perturb the Planck blackbody energy spectrum of the CMB photons, creating a spectral distortion.  If both photon number-changing processes, (e.g., double Compton scattering and Bremsstrahlung) and energy-changing processes (e.g., Compton scattering) occur efficiently, then these distortions are quickly wiped out, leaving no trace of the decay in the CMB energy spectrum besides a temperature shift (which is unobservable by spectral measurements, but may be detectable in the anisotropies via $N_{\rm eff}$).
However, should these processes be inefficient at the time of energy injection and remain inefficient until the present time, the spectral distortions they cause may freeze in and become observable.

A detailed review of CMB spectral distortions can be found in, e.g.,~\cite{Tashiro:2014pga,Chluba:2015bqa}.  As a rule of thumb, energy injections at redshifts $z \gtrsim 2 \times 10^6$ generally do not survive to be detected as spectral distortions, as the aforementioned photon number- and energy-changing processes are extremely effective at their erasure.  When the redshift drops below $z \sim 2 \times 10^6$, double
Compton and Bremsstrahlung begin to abate, while Compton scattering continues
to redistribute the photon energy with efficiency.  Under these conditions, exotic energy injection will generally result in a Bose-Einstein energy spectrum with a chemical potential $\mu$; such a deviation from the Planck spectrum is also called a $\mu$-distortion.

As the Universe evolves to $z  \lesssim 10^4$, Compton scattering too becomes inefficient; at these times, the primary effect of photon scattering on hot electrons is to shift the low-energy photons in the Rayleigh-Jeans part of the spectrum to the high-energy Wien tail (i.e., upscattering).  Thus, exotic energy injection at these times will not be redistributed to an equilibrium form, but instead results in a so-called $y$-distortion to the Planck spectrum whose shape is distinct from that of a chemical potential~$\mu$.  In other words, the exact shape of the spectral distortion contains in principle some amount of information on the time of the energy injection, with a pure $\mu$- and pure $y$-distortion representing respectively the early and the late extremes.   We note however that while a $\mu$-distortion is uniquely associated with energy injections in the early Universe, astrophysics at low redshifts, e.g., the Sunyaev-Zeldovich effect at $z \lesssim 20$, can also produce $y$-distortions that are typically much larger than what can be expected from distortions at pre-recombination times.   Although this does not prevent us from using null measurements of $y$-distortion to set constraints on early exotic energy injection scenarios, a positive measurement may not necessarily signal new particle physics.

Measurements of the CMB energy spectrum by the FIRAS instrument aboard COBE currently limits $\mu$-type and $y$-type distortions to $|\mu| \lesssim  9 \times 10^{-5}$ and
$|y|  \lesssim  1.5 \times  10^{-5}$ (95\% C.L.), respectively~\cite{Mather:1990tfx,Fixsen:1996nj}.  On the $(\tau,\epsilon_{\rm em})$-plane
in Fig.~\ref{Fig:gravEtau}, these limits are primarily responsible for constraining the $\tau \lesssim 10^{13}$~s section of the FIRAS+{\it Planck} exclusion region (shaded red).%
\footnote{The analysis of Ref.~\cite{Lucca:2019rxf}, on which Fig.~\ref{Fig:gravEtau} is based, in fact used the full energy spectrum measurement from FIRAS (i.e., not just derived limits on the $\mu$ and $y$ parameters) 
to put constraints on the $(\tau,f_{\rm frac})$-plane.}
A PRISM-like spectral measurement in the future could improve the sensitivity to $\delta \mu \sim 9 \times 10^{-10} (2 \sigma)$~\cite{Lucca:2019rxf,PRISM:2013fvg}, and is the main reason behind the improved constraints on $\epsilon_{\rm em}$ anticipated from future CMB experiments (shaded brown) at $\tau \lesssim 10^{12}$~s.


\subsection{CMB anisotropies}
\label{sec:aniso}

At $t \gtrsim 10^{12}$~s, electromagnetic energy injection into the cosmic plasma also begins to interfere with the atoms (mostly Hydrogen) in the Universe in an observable way via ionisation, excitation, and heating. This interference modifies directly the evolution of the free electron fraction---and hence how transparent the Universe is to photons---and has strong manifestations in the CMB temperature and polarisation anisotropies.
Energy injection around the time of CMB formation ($t \sim 10^{13}$~s) has a particularly strong impact, as it would delay recombination, thereby enhancing Silk damping and hence leading to a suppressed temperature-temperature angular power spectrum on small scales; it is to these timescales that the CMB anisotropies are maximally sensitive to energy injection.
Recombination itself is not affected if the energy injection occurs at a later time $t \gg 10^{-13}$~s.  Nonetheless, increased ionisation of the intergalactic medium between the epochs of recombination ($z \sim 1100$) and reionisation ($z \sim 10$) raises the optical depth, which manifests itself most prominently in a stronger E-polarisation signal at multipoles $\ell \sim 20$~\cite{Poulin:2016anj}.

On the $(\tau,\epsilon_{\rm em})$-plane shown in Fig.~\ref{Fig:gravEtau}, constraints from the {\it Planck} measurements of the CMB anisotropies are primarily responsible for $\tau \gtrsim 10^{13}$~s part of the FIRAS\texttt{+}\textit{Planck} exclusion region (shaded red).  In the future, the combination of LiteBIRD\texttt{+}CMB-S4\texttt{+}PRISM (shaded brown) can be expected to improve upon existing constraints.  However, as can be seen Fig.~\ref{Fig:gravEtau} and also discussed above in Sec.~\ref{sec:spectdist}, the largest improvement over current CMB constraints on  $\epsilon_{\rm em}$ pertains to shorter decay lifetimes $\tau \lesssim 10^{12}$~s and is due mainly to better spectral distortion measurements from a PRISM-like instrument; the improvement on the $\epsilon_{\rm em}$ constraints expected for longer lifetimes $\tau \gtrsim 10^{13}$~s from future CMB anisotropy measurements is relatively modest.

Lastly, we note again that our $\epsilon_{\rm em}$ constraints from the CMB have been mapped from the $f_{\rm frac}$ constraints of Ref.~\cite{Lucca:2019rxf}.  This mapping is direct, provided that the energy injection exceeds $\sim 1$~MeV and that the energy is deposited in the plasma instantly upon injection.


\subsection{Lyman-\texorpdfstring{$\alpha$}{alpha} forest}
\label{sec:lya}

The production of superWIMPs via NLSP-to-LSP decay is always accompanied by a finite momentum for the LSP, given at the production time $t_{\rm prod}$ by $p_{\rm LSP}(t_{\rm prod})= p_\star = (m^2_{\chi_1^0} - m_{\rm LSP}^2)/(2 m_{\chi_1^0})=\epsilon_{\rm em} m_{\chi_1^0}$, assuming decay at rest.  Universal expansion causes $p_{\rm LSP}$ to redshift subsequently as $p_{\rm LSP}(t)= p_\star R_{\rm prod}/R(t)$, where $R(t)$ is the scale factor and $R_{\rm prod}\equiv R(t_{\rm prod})$; at a later time $t$ the corresponding LSP velocity therefore reads
\begin{equation}
\begin{aligned}
\label{eq:vlsp}
v_{\rm LSP}(t) & = \frac{p_\star}{\sqrt{p_\star^2+ m^2_{\rm LSP} R^2(t)/R^2_{\rm prod}}}\\
&= \frac{1}{\sqrt{1+R^2(t)/v_0^2}},
\end{aligned}
\end{equation}
where
\begin{equation}
\begin{aligned}
v_0 & \equiv \epsilon_{\rm em} \frac{m_{\chi_1^0}}{m_{\rm LSP}}  R_{\rm prod} \\
& = \frac{\epsilon_{\rm em}}{\sqrt{1-2 \epsilon_{\rm em}}}  R_{\rm prod}
\end{aligned}
\end{equation}
is the present-day ($R(t_0)=1$) LSP velocity, under the assumption that $m_{\rm LSP} \gg p_{\rm LSP}(t_0)$ holds.

Because the NLSP-to-LSP decay is isotropic, the overall effect of a finite $v_{\rm LSP}$ is that of isotropic LSP free-streaming. Furthermore, because NLSP-to-LSP decay is a continuous process, together with a redshifting $p_{\rm LSP}$ we can expect a present-day comoving LSP number density $n_{\rm LSP}(t_0)$ to be composed of a distribution of particles in momentum space,  given approximately by 
\begin{equation}
\begin{aligned}
\label{eq:spectrum}
&n_{\rm LSP}(t_0) \simeq \frac{\Omega_{\rm DM} \rho_{\rm crit}}{m_{\chi_1^0}} \\
& \quad \times \Bigg\{ 2\int_0^{p_\star R_{\rm eq}}  {\rm d} \ln p \, \left(\frac{p}{p_\star R_\Gamma} \right)^2 
\exp\left[-\left(\frac{p}{p_\star R_\Gamma} \right)^2 \right]\\
&\quad +\frac{3}{2}\int_{p_\star R_{\rm eq}}^{p_\star}  {\rm d} \ln p \, \left(\frac{p}{p_\star R_\Gamma} \right)^{3/2} 
\exp\left[-\left(\frac{p}{p_\star R_\Gamma} \right)^{3/2} \right] \Bigg\},
\end{aligned}
\end{equation}
where the prefactor assumes the $\chi_1^0$  population was produced with the correct relic abundance, $R_\Gamma$ is the scale factor corresponding to NLSP lifetime $t=\tau \equiv 1/\Gamma$, and $R_{\rm eq}$ is the scale factor at matter-radiation equality.  See Appendix~\ref{sec:appendixA} for the derivation of Eq.~(\ref{eq:spectrum}).
Thus, phenomenologically, the LSP population today is dispersive and akin to a warm dark matter (WDM) with a present-day characteristic velocity given approximately by $v_0 \sim p_\star R_\Gamma/m_{\rm LSP}$.
This also means that limits on thermal WDM properties from small-scale fluctuation measurements such as the Lyman-$\alpha$ forest  can be reinterpreted to constrain properties of the LSP and the corresponding superWIMP parameter space.

To estimate these constraints, the critical quantity to compute is the comoving particle horizon (also called the free-streaming horizon) of the LSP population at a time~$t_{\rm obs}$ at which
the small-scale fluctuations are measured.  For a fixed production time~$t_{\rm prod}$, this is given by
\begin{equation}
\begin{aligned}
\label{eq:lambdaFS1}
\lambda_{\rm FS} (t_{\rm obs},t_{\rm prod}) & = \int_{t_{\rm prod}}^{t_{\rm obs}} \frac{{\rm d}t}{R(t)} v_{\rm LSP}(t)\\
& = \int^{R_{\rm obs}}_{R_{\rm prod}} \frac{{\rm d}R}{R^2 H(R)} v_{\rm LSP}(R),
\\
\end{aligned}
\end{equation}
where $t_{\rm obs}$ corresponds typically to a low redshift of $z  \sim 2$, $t_{\rm prod}$ to some time prior to matter-radiation equality, and $H\equiv (1/R)({\rm d} R/{\rm d} t)$ is the Hubble expansion rate.
 Because NLSP decay is a continuous process, in principle we must calculate $\lambda_{\rm FS}$ for all possible production times~$t_{\rm prod}$ and then average it over the momentum distribution of Eq.~(\ref{eq:spectrum}).  For simplicity, however, we assume all LSPs to be produced at $t_{\rm prod}=1/\Gamma$; this is a reasonable approximation given that, as shown in Eq.~(\ref{eq:spectrum}), LSPs produced at $t_{\rm prod}\sim 1/\Gamma$---whose characteristic momentum is $\sim p_\star R_\Gamma$---in any case dominate the LSP distribution today. Then, using Eq.~(\ref{eq:vlsp}) and setting $t_{\rm prod} = 1/\Gamma$, Eq.~(\ref{eq:lambdaFS1}) can now be rewritten as
\begin{equation}
\begin{aligned}
& \lambda_{\rm FS}(R_{\rm obs}, R_\Gamma)  \\
 & \qquad = \frac{1}{H_0}\sqrt{\frac{R_{\rm eq}}{\Omega_{\rm m}}} \int_{R_\Gamma/R_{\rm eq}}^{R_{\rm obs}/R_{\rm eq}}  \frac{{\rm d} y}{\sqrt{(1+y)[1+ (R_{\rm eq}/v_0)^2 y^2] }} \\
 & \qquad   \simeq 91 \, h^{-1}{\rm Mpc}\;  \sqrt{\left(\frac{3401}{1+z_{\rm eq}} \right) \left(\frac{0.32}{\Omega_{\rm m}} \right)} \\
 & \qquad \quad \times \int_{R_\Gamma/R_{\rm eq}}^{R_{\rm obs}/R_{\rm eq}}  \frac{{\rm d} y}{\sqrt{(1+y)[1+ (R_{\rm eq}/v_0)^2 y^2] }},
\end{aligned}
\end{equation}
valid across the radiation and matter domination epochs, where the subscript ``eq'' denotes matter-radiation equality, $\Omega_{\rm m}$ is the present-day reduced total matter density, and $h$ is the reduced Hubble parameter defined via $H_0 = 100 \, h~{\rm km} \, {\rm s}^{-1} \, {\rm Mpc}^{-1}$.

The typical Ly$\alpha$ WDM bound in the literature is presented as a lower limit on the WDM mass $m_{\rm WDM}$, assuming that the WDM constitutes the entirety of the dark matter abundance of the Universe and that the WDM population follows a relativistic Fermi-Dirac distribution, with a temperature linked to the dark matter abundance, i.e.,
\begin{equation}
\label{eq:thermalWDM}
   \Omega_{\rm WDM}  h^2= \left(\frac{T_{\rm WDM}}{T_\nu}\right)^3 \left(\frac{m_{\rm WDM}}{94~{\rm eV}} \right),
\end{equation}
where $T_\nu$ is the temperature of the neutrino background ($T_{\nu,0} = 1.95$~K).  Fixing $\Omega_{\rm WDM} h^2 = 0.12$ and using the current best limit $m_{\rm WDM} \gtrsim 5.3$~keV (95\% C.I.)~\cite{Irsic:2017ixq}, we find an upper bound on the present-day WDM temperature of $T_{{\rm WDM},0} \lesssim 2.2 \times 10^{-5}$~eV, or equivalently, a bound on the present-day average WDM velocity of $v_{\rm WDM,0} \simeq 3\, T_x/m_x \lesssim  1.2 \times 10^{-8}$.  Following the steps outlined above and letting $R_{\rm prod} \to 0$, it is straightforward to show that the equivalent upper limit on the comoving WDM free-streaming horizon at $z_{\rm obs}=2$ is
\begin{equation}
\label{eq:lambdaWDM}
\lambda_{\rm FS}^{\rm WDM} (z_{\rm obs}=2) \lesssim 0.045 \, h^{-1} {\rm Mpc}  \; \sqrt{\left(\frac{3401}{1+z_{\rm eq}} \right) \left(\frac{0.32}{\Omega_{\rm m}} \right)}.
\end{equation}
This limit can in principle serve as an upper bound on $\lambda_{\rm FS}(z_{\rm obs}=2,z_\Gamma)$ for those superWIMP scenarios wherein the LSP explains {\it all} of the observed dark matter abundance.%
\footnote{Constraints on the free-streaming properties of the dark matter can also be derived from the distribution of Milky Way satellite galaxies.  Observations currently limit the thermal WDM mass to $m_{\rm WDM} \gtrsim 6.5$~keV~(95\% C.I.)~\cite{DES:2020fxi} assuming $f_{\rm WDM}=1$, marginally better than the Ly$\alpha$ bound of Ref.~\cite{Irsic:2017ixq}.  We therefore do not consider Milky Way constraints here.}

\begin{table}[t]
\caption{Upper limit on the free-streaming horizon $\lambda_{\rm FS}^{\rm WDM}$ as a function of the WDM fraction $f_{\rm WDM}$.  These limits have been mapped from the two-dimensional 68\%-credible contours in $(f_{\rm WDM},m_{\rm WDM})$-plane in Fig.~12 of Ref.~\cite{Boyarsky:2008xj}, which dates from 2009.  In the comparable case of $f_{\rm WDM}=1$, we note that the limit $m_{\rm WDM} \gtrsim 5.3$~keV (95\%~C.I.) quoted in the text comes from a more recent 2017 analysis~\cite{Irsic:2017ixq} and translates to $\lambda_{\rm FS}^{\rm WDM} \lesssim 0.045\, h^{-1}$Mpc (see also Eq.~(\ref{eq:lambdaWDM})), in contrast to $\lambda_{\rm FS}^{\rm WDM} \lesssim 0.0708\, h^{-1}$Mpc tabulated below.  The Ly$\alpha$ limits used in our analysis are therefore quite conservative.\label{tab:lya}}
\begin{ruledtabular}
\begin{tabular}{cc}
$f_{\rm WDM}$ & Upper limit on $\lambda_{\rm FS}^{\rm WDM}$ [$h^{-1}$Mpc] \\
\hline
0.15     & 0.323 \\
0.2     & 0.272 \\
0.3 & 0.208 \\
0.4 & 0.156 \\
0.5 & 0.130 \\
0.6 & 0.113 \\
0.7 & 0.102 \\
0.8 & 0.0938 \\
0.9 & 0.0813 \\
1.0 & 0.0708
\end{tabular}
\end{ruledtabular}
\end{table}

A more versatile analysis of the Ly$\alpha$ data could however also consider the possibility of a mixed cold+warm dark matter cosmology and vary as part of the fitting procedure the fraction of the total dark matter in the form of WDM, $f_{\rm WDM}  \equiv \Omega_{\rm WDM}/\Omega_{\rm DM}$.  Reference~\cite{Boyarsky:2008xj} has provided such an analysis and presented the outcome as two-dimensional constraints in the $(f_{\rm WDM},m_{\rm WDM})$-plane and in the $(f_{\rm WDM},v_0)$-plane (see Fig.~12 of~\cite{Boyarsky:2008xj}).  We have translated these constraints to constraints in the  $(f_{\rm WDM},\lambda_{\rm FS}^{\rm WDM})$-plane.  See a representative set in Table~\ref{tab:lya}.  Observe that the upper limit on $\lambda_{\rm FS}^{\rm WDM}$ deteriorates as we decrease the WDM fraction $f_{\rm WDM}$; at $f_{\rm WDM} \lesssim 0.15$, no limit can be set on $\lambda_{\rm FS}^{\rm WDM}$.
In our analysis of superWIMPs, we take the WDM fraction to be $f_{\rm WDM} = m_{\rm LSP}/m_{\rm NLSP}$ and assume the remaining cold dark matter to be explained by some other physics.  We apply the Ly$\alpha$ constraints only to those cases where production takes place before matter-radiation equality, i.e., $1/\Gamma \leq t_{\rm eq}$, because using $\lambda_{\rm FS}$ as a proxy for small-scale suppression is likely unreliable if the bulk of the NLSP decay happens deep in matter domination.%
\footnote{Since we must always compute $v_0$, the possibility also exists to constrain superWIMP scenarios using the thermal WDM limit on $v_{{\rm WDM},0}$, instead of the more complicated $\lambda_{\rm FS}^{\rm WDM}$.  Indeed, the limits in the superWIMP parameter space do turn out to be quite similar if we restrict $t_{\rm prod}$ to the radiation-domination epoch.  Conceptually though, we note that $v_0$ describes only the instantaneous free-streaming behaviour of the LSP population, while $\lambda_{\rm FS}$ is able to capture a more complete free-streaming history; the difference between the two measures becomes more marked if LSP production takes place during matter domination.  We emphasise however that neither measure is likely accurate if NLSP-to-LSP decay happens deep in the matter domination epoch, as the growth histories of the density perturbations in such scenarios---especially as the growth enters the  nonlinear regime---deviate too strongly from the conventional thermal WDM scenario.\label{foot:fs}}


\subsection{Collider constraints}
\label{sec:collider}

We assume from the outset that the neutralino 
is not ruled out by conventional jets/leptons + missing energy searches. In principle, within the scope of specific mass spectra, part of the parameter space can indeed be ruled out; however, this requires a larger global fit within specific simplified or full SUSY models. The GAMBIT collaboration has performed a global fit of the pMSSM  within a reduced 7-parameter space---the relevant parameters being the trilinear couplings, Higgsino mass parameters, diagonal sfermion masses, and $\tan\beta$---taking into account collider, direct detection and relic density observables~\cite{GAMBIT:2017zdo}. They conclude that a large volume of the pMSSM parameter space remains unconstrained, with the neutralino mass ranging from the $Z$/$H$ funnel region to the mulit-TeV scale. In an extended set-up that includes gravitinos, for a gravitino mass fixed at $m_{\tilde{G}}=1$~eV, best-fit points that take into account collider searches for the rest of the electroweak gaugino sector indicate that both light and heavy neutralinos remain unconstrained within a variety of simplified-model scenarios~\cite{GAMBIT:2023yih}. Scenarios like split SUSY models~\cite{Arkani-Hamed:2004ymt,Giudice:2004tc} also predict mass spectra that are unconstrained by LHC searches.

Bearing the above in mind,
we summarise in this subsection collider constraints on the gravitino and the axino LSP originating from neutralino decay.  
The neutralino proper decay length to gravitino can be expressed following Eq.~(\ref{eq:GravitinoDecayWidth2}) as a function of the fractional energy $\epsilon_{\rm em}$ and the neutralino mass $m_{\chi_{1}^{0}}$:
\begin{equation}
L= c\tau\simeq  1.4 \times 10^{22}~{\rm m}\,  \frac{1-2\epsilon_{\rm em} }{\epsilon_{\rm em}^{3}  (2-3\epsilon_{\rm em}) } \left(\frac{\rm GeV }{m_{\chi_{1}^{0}}}\right)^{3} .
\end{equation}
Collider searches---including searches of prompt decays that occur at the interaction vertex and are hence sensitive to photons plus missing energy signatures, and LLP searches sensitive to delayed decays---probe proper decay lengths of about 100~m.  Thus, in order for colliders to be sensitive to the gravitino superWIMP scenario, it is {\it a priori} clear that a large mass hierarchy must exist between the neutralino and the gravitino.

The LEP experiment has placed a lower bound on the gravitino mass from the process $e^{+}e^{-} \to \tilde{G}\tilde{G}\gamma$ of $m_{\tilde{G}}\gtrsim 1.09 \times 10^{-5}$ eV~\cite{DELPHI:2003dlq}. Furthermore, under the assumption that the  rest of the SUSY spectrum is decoupled apart from the selectron and the neutralino $\chi_{1}^{0}$, the LEP searches exclude a neutralino mass of up to $m_{\chi_{1}^{0}}\simeq$ 200 GeV for a  gravitino mass of $m_{\tilde{G}}\lesssim 10^{-5}$ eV. At the LHC, gravitino searches have been conducted within the context of  Gauge Mediated Supersymmetry (GMSB) breaking models~\cite{Giudice:1998bp}. These searches look for displaced photons assuming a SUSY topology that yields the neutralino NLSP.  Assuming a decay channel with maximal production cross-section in the $pp \to \tilde{q}\tilde{q}\to qq\chi_{1}^{0}\chi_{1}^{0}$ followed by the displaced photon signature of $\chi_{1}^{0}\to \tilde{G}\gamma$, the ATLAS experiment rules out neutralino  masses in the range $\sim 100-400$~GeV for $c\tau \sim 10-10^{4}$~cm~\cite{ATLAS:2014kbb}. The latest CMS result \cite{CMS:2019ulr} at 13 TeV with 78 $\rm fb^{-1}$ luminosity excludes within these scenarios a neutralino mass in the range $m_{\chi_1^0}\sim 200-550$~GeV, for $c\tau \sim 10-10^{4}$~cm. We will use these experimental bounds for illustrative purposes, but emphasise that they are model-dependent.

Similar considerations apply to the axino superWIMP, in which case we have an additional degree of freedom in the neutralino decay width, namely the axion decay constant $f_a'\equiv f_a/N$. The proper decay length for the axino follows simply from Eq.~(\ref{eq:decaywidthaxino}): 
\begin{equation}
   L= c\tau\simeq  14.15~{\rm m} ~ \epsilon_{\rm em}^{-3} \left(\frac{f_{a}^{'}}{10^{8}~\rm GeV}\right)^{2}\left(\rm \frac{100~\rm GeV}{m_{\chi_{1}^{0}}}\right)^{3}.
\end{equation}
As with the gravitino, collider constraints on the axino superWIMP scenario depend in general on the model specifics.  Independently of the specifics, however, in order for colliders to be sensitive to the parameter space, the decay length should be $\lesssim \mathcal{O} (100)$~m. Within the context of specific models, estimates have been made on the capability of the LHC to probe axinos from neutralino decays in prompt and LLP searches~\cite{Co:2016fln}. In this work we will reinterpret existing LLP search results to put limits on this decay process.

Finally, since we are dealing with long-lived neutralinos, if they are light (GeV/sub-GeV), there is potentially sensitivity at fixed-target experiments such as CHARM~\cite{CHARM:1983ayi,Gninenko:2012eq}, NA62~\cite{Dobrich:2018ezn}, NOMAD~\cite{NOMAD:2001eyx}, SHiP~\cite{SHiP:2015vad} and SEAQuest~\cite{PhysRevD.98.035011}, as well as at forward physics facility experiments like FASER~\cite{Feng:2022inv}. 

The production of neutralinos in fixed-target experiments depend on the specifics of the model. Typically in such experiments, a beam of particles with energies ranging from $\sim 100$~GeV (SEAQuest) to $\sim 450$~GeV (SHiP, NOMAD, NA62) collides with the target, thereby producing hadrons and weakly-interacting particles that could be captured in a far detector.  If the rest of the SUSY spectrum is decoupled and heavy, neutralino production proceeds primarily from the decay of pseudo-scalar and vector mesons. These neutralinos then decay to gravitinos/axinos with decay widths given by Eqs.~(\ref{eq:GravitinoDecayWidth}) and~(\ref{eq:decaywidthaxino}).  
 Thus, the number of long-lived neutralino decays within a fixed-target experiment depends on the decay lifetime, the Lorentz boost factor, and the energy spectrum of the mesons specific to the experiment. While a full analysis for each experiment is beyond the scope of this work, a crude estimate suggests that the NOMAD experiment can exclude up to $m_{\chi_{1}^{0}}\simeq 300$~MeV for a fixed $m_{\tilde{a}}\simeq 20$ MeV, assuming $f_{a}\simeq 10^{3}$ GeV. For the same fixed value of $m_{\tilde{a}}$, a future experiment like SHiP can rule out $f_{a}=10^{4}$~GeV for $m_{\chi_{1}^{0}}\simeq 300$~MeV\footnote{Also see detailed estimates provided by \cite{Choi:2019pos}, with which we agree.}.

As for FASER, a recent assessment of its feasibility to probe light axinos and gravitinos found that, for \mbox{$m_{\tilde{a}}\simeq 10$}~MeV, it is possible to rule out \mbox{$f_{a} \simeq 10^{2}-10^{3}$}~GeV for $m_{\chi_{1}^{0}}\simeq 300$~MeV~\cite{Jodlowski:2023fvz}.


\section{Impact of Constraints on the superWIMP parameter space}
\label{sec:results}

Having discussed the relevant cosmological and collider probes of superWIMPs, we are now in a position to present the constraints on the gravitino and the axino superWIMP parameter spaces. These are shown in Figs.~\ref{Fig:gravmass1} and~\ref{Fig:axinomass} in the $(m_{\chi_1^0},m_{\rm LSP})$- and $(m_{\chi_1^0},\epsilon_{\rm em})$-projections, highlighting respectively the hierarchical ($m_{\chi_1^0} \gg m_{\rm LSP}$) and degenerate ($m_{\chi_1^0} \simeq m_{\rm LSP})$ regions (recall that \mbox{$2\epsilon_{\rm em} = \Delta m^2/m_{\chi_1^0}^2$}, where $\Delta m^2 \equiv m_{\chi_1^0}^2 - m_{\rm LSP}^2$ is the NLSP-LSP squared-mass gap).
Detailed discussions of these results follow below.


\subsection{Gravitino}
\label{sec:gravitinoconstraints}

\begin{figure*}[t]
    \centering
\includegraphics[width=.48\textwidth]{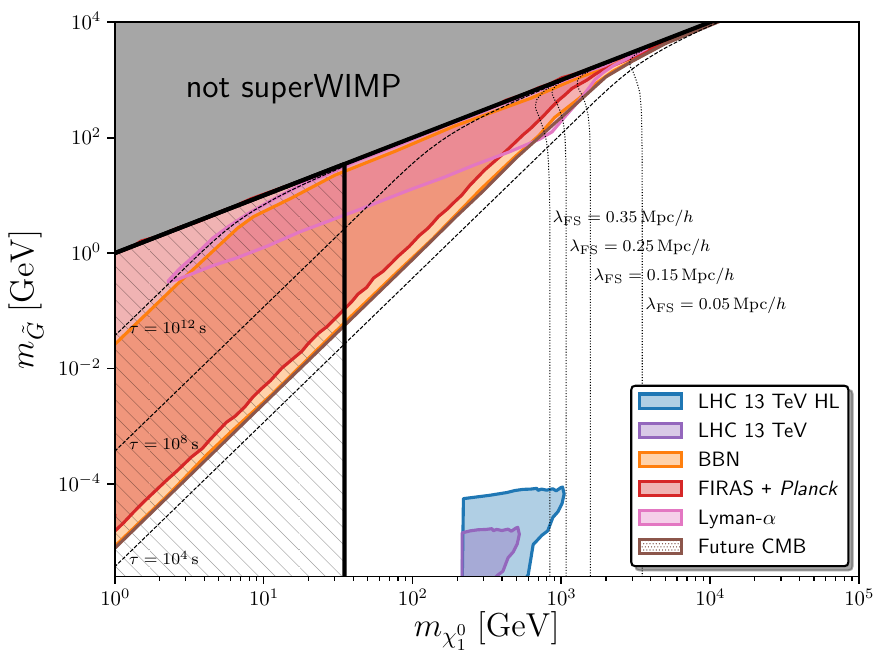}
\includegraphics[width=.48\textwidth]{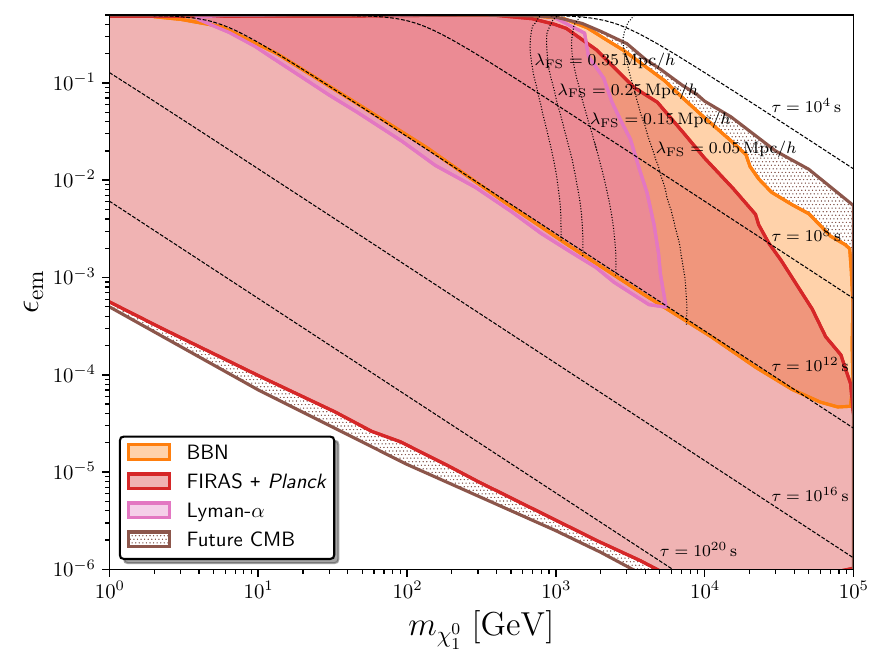}
    \caption{{\it Left}: Current and projected constraints on the gravitino superWIMP parameter space in the $(m_{\chi_1^0},m_{\tilde{G}})$-plane.  Exclusion regions labelled BBN, FIRAS\texttt{+}{\it Planck}, and FutureCMB are based on the energy injection considerations of Fig.~\ref{Fig:gravEtau}.  The Lyman-$\alpha$ constraints are derived from upper limits on the WDM free-streaming horizon as a function of the WDM fraction in Table~\ref{tab:lya}. 
    Collider constraints from the LHC exclude the region shaded in purple, while  projected constraints from HL-LHC are shown in blue.
    The region marked ``not superWIMP'' does not satisfy the superWIMP condition $m_{\chi_1^0} > m_{\tilde{G}}$, while  the maximum value on the horizontal axis, $m_{\chi_1^0}=100$~TeV, corresponds to the unitarity limit (see Footnote~\ref{footnote:unitary}).
 The hatched region indicates where a neutralino abundance will overclose the Universe in the context of the freeze-out mechanism given current collider constraints on the neutralino couplings~\cite{Barman:2017swy}; we do not however enforce this constraint, since any one of non-thermal production mechanism, late entropy injection, or modified cosmological histories could in principle weaken the exclusion limit.   
       For clarity, we also show several lines of constant neutralino lifetime (black short dashed lines) and constant gravitino free-streaming horizon (black dotted lines).
    {\it Right}: Similar to the left panel, but in the $(m_{\chi_1^0},\epsilon_{\rm em})$-plane and without LHC constraints. 
    }
    \label{Fig:gravmass1}
\end{figure*}

Fig.~\ref{Fig:gravmass1} shows the current and projected cosmological and collider constraints on the gravitino superWIMP parameter space in the $(m_{\chi_1^0},m_{\tilde{G}})$-plane (left panel) and the 
$(m_{\chi_1^0},\epsilon_{\rm em})$-plane (right panel).  A quick glance at both reveals that a large parameter region from the hierarchical to the degenerate limit is strongly constrained by a plethora of cosmological observations.  
In particular, we observe that for a neutralino mass fixed at $m_{\chi_{1}^{0}}=100$~GeV, the totality of cosmological observations rules out gravitino masess lying in the range 
\begin{equation}
\label{eq:gravEx1}
0.8 \lesssim m_{\tilde{G}}/{\rm GeV} \lesssim 99.998,
\end{equation}
while for $m_{\chi_{1}^{0}}=1$~TeV the exclusion region lies in the  range 
\begin{equation}
\label{eq:gravEx2}
270 \lesssim m_{\tilde{G}}/{\rm GeV} \lesssim  999.997.
\end{equation}
As a rule of thumb, the lower end of these exclusion regions in the gravitino mass $m_{\tilde{G}}$ is driven by how {\it short} a neutralino lifetime $\tau$ a particular cosmological observation can probe with some precision, and is currently dominated by energy injection constraints from observations of the primordial $^2$H and $^3$He abundances.  This limit can be seen most clearly in the left panel of Fig.~\ref{Fig:gravmass1} represented by the lower edge of the orange-shaded area at $m_{\chi_1^0} \lesssim 1$~TeV,
and in the right panel by the upper edge of the orange-shaded area at $m_{\chi_1^0} \gtrsim 1$~TeV.  A future CMB spectral measurement by a PRISM-like instrument will contribute to tightening the lower limit on $m_{\tilde{G}}$ primarily in the degenerate ($m_{\chi_1^0} \simeq m_{\tilde{G}}$) region at $m_{\chi_1^0}\gtrsim 1$~TeV (right panel, upper edge of the brown-shaded area).

On the other hand, the upper end of the exclusion ranges~(\ref{eq:gravEx1}) and (\ref{eq:gravEx2}) is mainly determined by how {\it long} a decay lifetime an observation can probe, and in this regard, it is the energy injection limits from the {\it Planck} CMB anisotropy measurements that dominate the current constraints.  The {\it Planck} CMB anistropy limit is most evident in the right panel of Fig.~\ref{Fig:gravmass1} (lower edge of the red-shaded area), and extends to lifetimes well in excess of $\tau \sim 10^{20}$~s, i.e., longer than the lifetime of the Universe thus far ($t\sim 4 \times 10^{17}$~s).  Improvement from a future CMB anisotropy measurement by LiteBIRD\texttt{+}CMB-S4 in this region will however likely be marginal (right panel, lower edge of the brown-shaded area), as the reach in decay lifetime of these future observations is unlikely to outperform the {\it status quo} by more than a factor of a few (see also Fig.~\ref{Fig:gravEtau}).

It is also interesting to consider what limits Ly$\alpha$ observations impose on the gravitino superWIMP parameter space; the corresponding exclusion region is shaded pink in both the left and right panels of Fig.~\ref{Fig:gravmass1}.  At face value the Ly$\alpha$ bounds do not appear to add much to the energy injection constraints already discussed above in either  the $(m_{\chi_1^0},m_{\tilde{G}})$- or $(m_{\chi_1^0},\epsilon_{\rm em})$-plane.  However, while BBN and CMB observations probe the electromagnetic energy injected into the cosmic plasma, Ly$\alpha$ is sensitive to the free-streaming properties of the gravitino itself and therefore offers a somewhat different perspective on the superWIMP scenario.  It also provides a useful constraint on those scenarios not considered in this work wherein the NLSP does not decay electromagnetically (e.g., decay into neutrinos).
We devote several paragraphs below to explain the essential features of the Ly$\alpha$ constraint on the gravitino superWIMP parameter space.

Observe first in the left panel of Fig.~\ref{Fig:gravmass1} that Ly$\alpha$ rules out a substantial region around neutralino masses of ${\cal O}(1)$~GeV to ${\cal O}(1)$~TeV in the $(m_{\chi_1^0},m_{\tilde{G}})$-plane, and, like the energy injection constraints, provides an upper limit on viable values of~$m_{\tilde{G}}$, albeit a much weaker one.  This may at first glance appear counter-intuitive to the common understanding that Ly$\alpha$ observations limit the particle masses of free-streaming dark matter from below, to $m_{\rm WDM} \gtrsim {\cal O}(1)$~keV.  To reconcile these seemingly conflicting results, note first of all that thermal WDM mass constraints in the literature typically carry the assumption that the WDM makes up all of the dark matter content of the Universe.  On the other hand, the upper limit on $m_{\tilde{G}}$ in Fig.~\ref{Fig:gravmass1} arises from the fact that once the WDM fraction---defined here as $f_{\rm WDM}= m_{\tilde{G}}/m_{\chi_1^0}$ (see Sec.~\ref{sec:lya})---drops below $f_{\rm WDM} <0.15$, the free-streaming properties of the gravitino LSP become unconstrainable by current observations (see also Table~\ref{tab:lya}).  In fact, this upper limit on $m_{\tilde{G}}$ parallels cosmological bounds on the absolute mass scale of Standard-Model neutrinos, where extremely small masses (and hence very large free-streaming scales) cannot be constrained because the energy density low-mass neutrinos contribute to the total dark matter content is
too minute for their free-streaming effects to be observable.

Secondly, free-streaming is a kinematic effect dependent only on the characteristic {\it velocity} of the WDM in question.  Its use as tool to constrain WDM masses to ${\cal O}(1)$~keV masses and above rests strongly on the assumption that the WDM has been produced via scattering processes with the SM thermal bath, such that the WDM population inherits the characteristic momentum (or temperature) of the bath {\it even} in the event that thermalisation is incomplete.
In the case of production via NLSP decay, however, the characteristic momentum inherited by the LSP at production, $p_\star= \epsilon_{\rm em} m_{\chi_1^0}$, is unrelated to the properties of the SM thermal bath; rather, it is determined by the NLSP mass---which, in our scenarios, is typically orders of magnitude larger than the SM bath temperature at the time of decay---and the NLSP-LSP mass gap. Consequently, the LSP masses constrained by velocity-based free-streaming arguments also fall in a range of orders of magnitude above na\"{\i}ve expectations.

Turning our attention now to the Ly$\alpha$ limits displayed in the right panel of Fig.~\ref{Fig:gravmass1}, we see immediately that there is in fact also a lower limit on viable values of $m_{\tilde{G}}$, manifesting in the $(m_{\chi_1^0},\epsilon_{\rm em})$-plane  
as a lower limit on viable values of $\epsilon_{\rm em}$ at $m_{\chi_1^0} \gtrsim 1$~TeV (right edge of the pink-shaded region). To explain this limit, recall that in this plot, the energy injection is generally tiny, i.e., $\epsilon_{\rm em} \ll 0.5$, such that the WDM fraction is saturated at $f_{\rm WDM} = 1$ over the bulk of the displayed parameter region.  Thus, the same $f_{\rm WDM} = 1$ free-streaming horizon limit ($\lambda_{\rm FS} \lesssim 0.0708\, h^{-1}$Mpc from Table~\ref{tab:lya}) applies (almost) everywhere.  Indeed, the right edge of the Ly$\alpha$ exclusion region aligns closely with the contours of constant free-streaming horizon $\lambda_{\rm FS}$ in the right panel of Fig.~\ref{Fig:gravmass1} as expected.  In contrast, the same edge in the $(m_{\chi_1^0},m_{\tilde{G}})$-plane in the left panel shows no such alignment; rather, the edge shifts to higher values of $\lambda_{\rm FS}$ as we decrease $m_{\tilde{G}}$, reflecting the weakening of the Ly$\alpha$ limit on $\lambda_{\rm FS}$ with decreasing $f_{\rm WDM}$ evident in Table~\ref{tab:lya}.

Finally, we remark that collider experiments constrain only a very tiny region of the gravitino superWIMP parameter space, in a spot where an extremely large hierarchy exists between $m_{\chi_1^0}$ and $m_{\tilde{G}}$ so as to produce a detectable LLP signal (left panel of Fig.~\ref{Fig:gravmass1}, purple-shaded region).  Specifically, reinterpreting the ATLAS and CMS limits from Refs.~\cite{ATLAS:2014kbb,CMS:2019ulr}, which are presented on the $(m_{\chi_1^0},\tau)$-plane with GMSB SPS8 benchmarks as the reference model, we find that 
for neutralino masses up to $m_{\chi_1^0} \simeq 400$~GeV, current LHC data exclude gravitino masses up to $m_{\tilde{G}} \simeq 10^{-5}$~GeV.   As discussed in Sec.~\ref{sec:collider}, LLP search limits are model-dependent: thus the LHC exclusion region presented in Fig.~\ref{Fig:gravmass1} can move (within small confines) depending on the model specifics of the search analysis.   

Note also that our LHC exclusion region has a vertical cut-off at $m_{\chi_1^0} \simeq 200$~GeV; this follows simply from the fact that Refs.~\cite{ATLAS:2014kbb,CMS:2019ulr} had chosen to focus on a specific region 
of parameter space in their analyses.  It is not inconceivable that current LHC data could constrain also some parts of the parameter region to the left of the cut-off; it is however extremely unlikely that the exclusion region extends all the way to $m_{\chi_1^0} =0$,  as the search would become background-limited. 

For completeness we estimate also the sensitivity of the high-luminosity LHC (HL-LHC) run, by simply rescaling the current LHC constraint to $3000~{\rm fb}^{-1}$ integrated luminosity (left panel, blue-shaded region). We find that HL-LHC can improve the reach in gravitino masses by at most an order of magnitude to $m_{\tilde{G}}\simeq 10^{-4}$ GeV for neutralino masses up to $m_{\chi_1^0} \simeq 1$~TeV.


\subsection{Axino}

\begin{figure*}[t]
    \centering
\includegraphics[width=.48\textwidth]{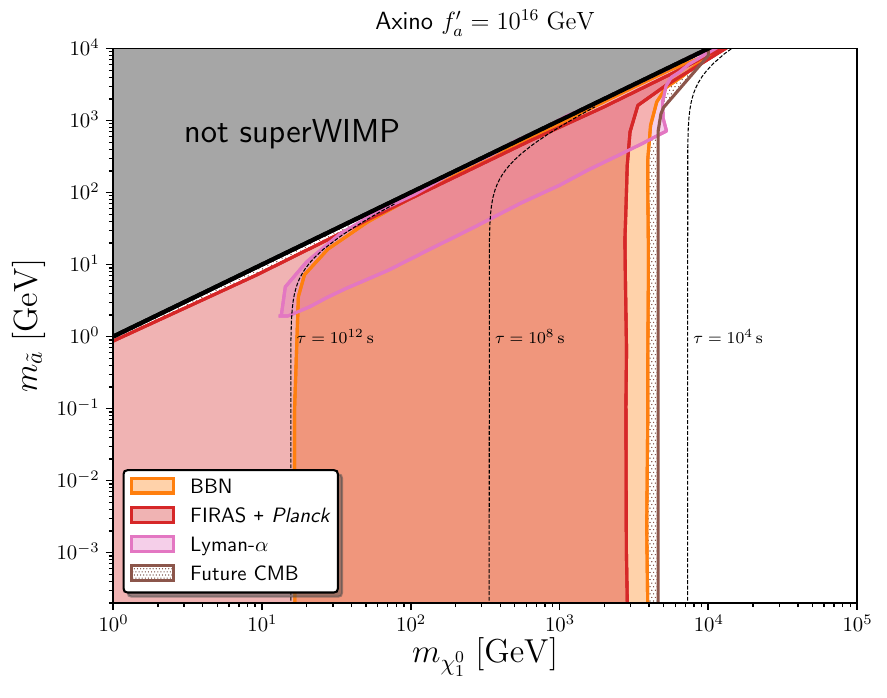}
\includegraphics[width=.48\textwidth]{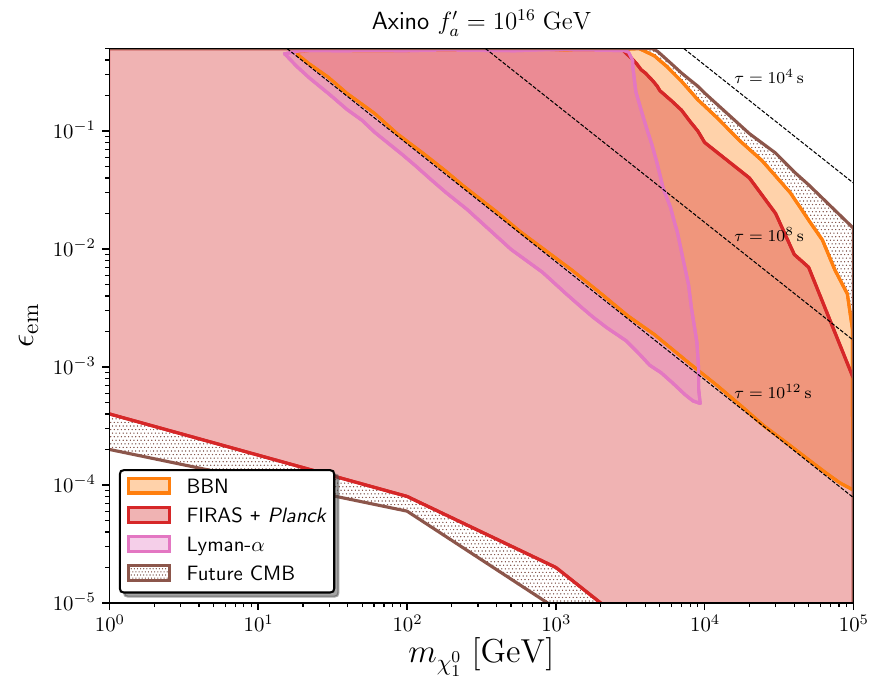}
\includegraphics[width=.48\textwidth]{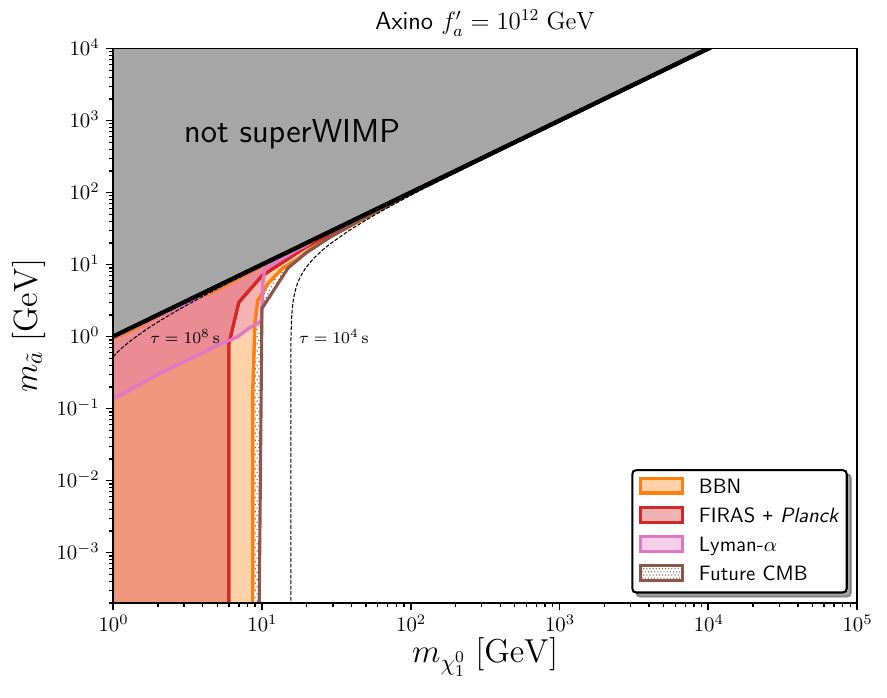}
\includegraphics[width=.48\textwidth]{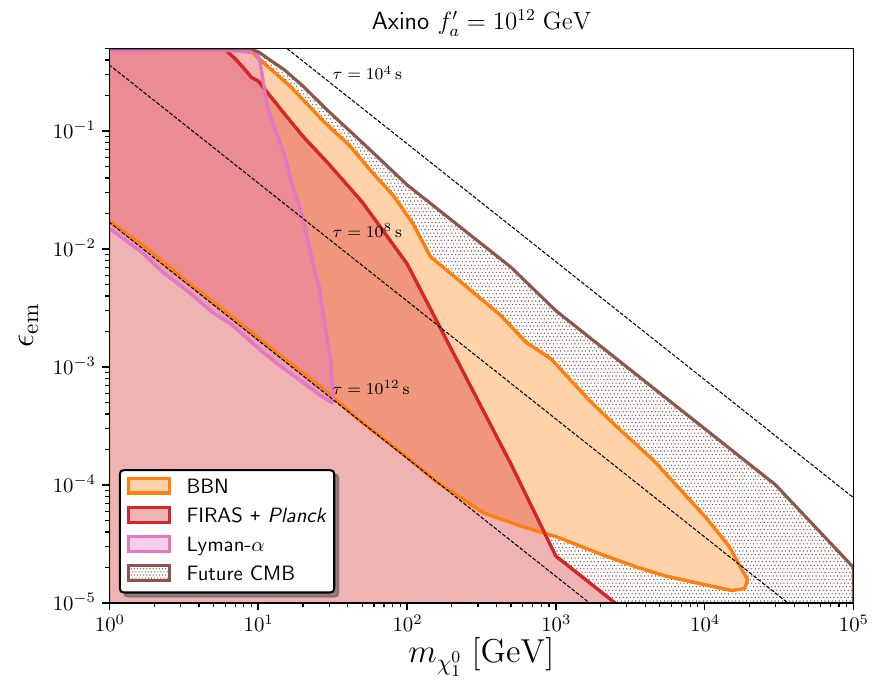}
\includegraphics[width=.48\textwidth]{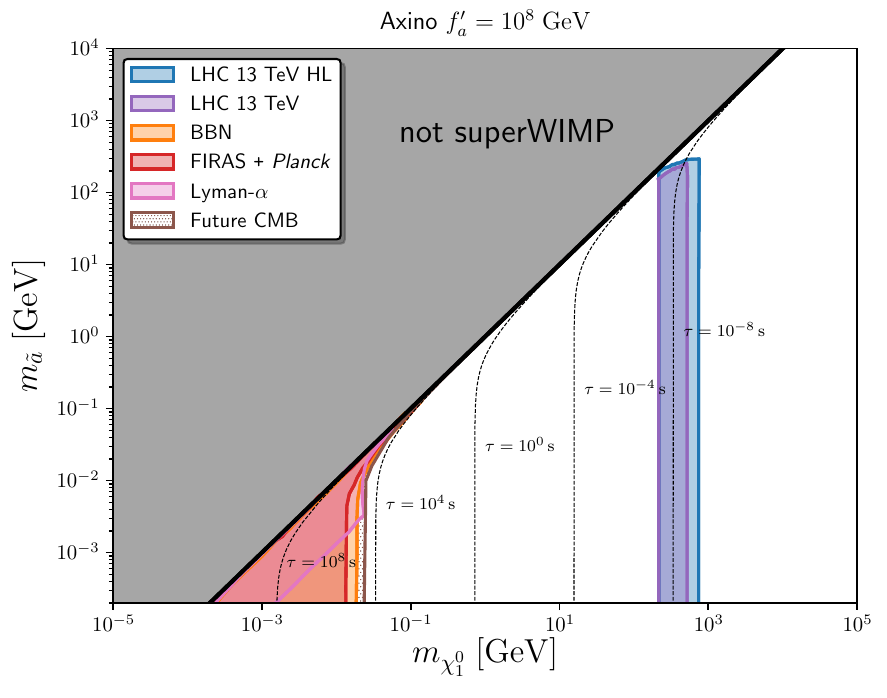}
\includegraphics[width=.48\textwidth]{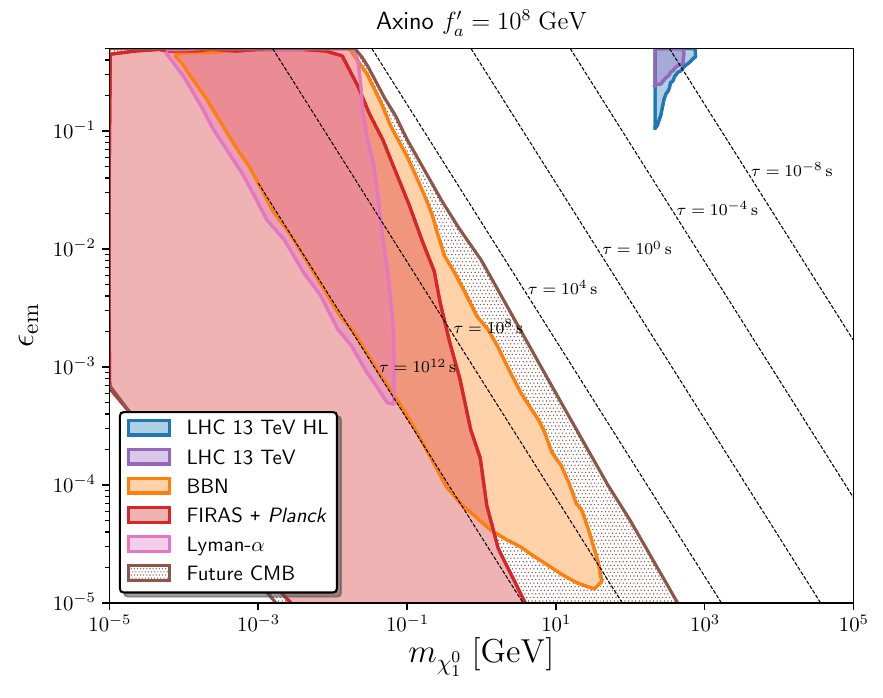}
    \caption{Same as Fig.~\ref{Fig:gravmass1}, but for the axino superWIMP assuming three different values of the axion decay constant.   We have however omitted plotting the lines of constant $\lambda_{\rm FS}$, as well as the exclusion region at $m_{\chi_1^0}\lesssim 34$~GeV, which we in any case do not enforce.}
    \label{Fig:axinomass}
\end{figure*}

Analogously to Fig.~\ref{Fig:gravmass1}, Fig. \ref{Fig:axinomass} shows the current and projected cosmological and collider constraints on the axino superWIMP parameter space in the $(m_{\chi_1^0},m_{\tilde{a}})$-plane (left panels) and the 
$(m_{\chi_1^0},\epsilon_{\rm em})$-plane (right panels), assuming three different values of the axion decay constant~$f_a'\equiv f_a/N \in [10^8, 10^{12}, 10^{16}]$~GeV.      As in Fig.~\ref{Fig:gravmass1}, overclosure excludes in principle $m_{\chi_1^0} \lesssim 34$~GeV if the neutralino is produced via thermal freeze-out; we have however omitted plotting this constraint in Fig.~\ref{Fig:axinomass} since it can be easily circumvented by other production mechanisms and we in any case do not enforce it.

As can be gleaned from the neutralino-to-axino decay width~(\ref{eq:decaywidthaxino}), with $\epsilon_{\rm em}$ held fixed, the decay lifetime scales as $\tau \propto {f_a'}^2/m_{\chi_1^0}^3$.  Thus, in both the $(m_{\chi_1^0},m_{\tilde{a}})$- and $(m_{\chi_1^0},\epsilon_{\rm em})$ planes, we generally expect experimental constraints (cosmological and collider) to apply to larger neutralino mass values as we crank up $f_a'$.  Indeed, as is evident in the left panels of Fig.~\ref{Fig:axinomass}, the right edges of the cosmological exclusion regions (BBN, \mbox{FIRAS}\texttt{+}{\it Planck}, and FutureCMB) mostly align with the lifetime reach of the observations at the short end.  Where the neutralino and axino masses are strongly hierarchical (so that $\epsilon_{\rm em} \simeq 0.5$), the decay lifetime also becomes independent of $m_{\tilde{a}}$.  Consequently, 
the points at which the edges of the exclusion regions intersect the horizontal $m_{\chi_1^0}$-axis 
also scale  as ${f_a'}^{2/3}$.  As in the case of the gravitino superWIMP, the most stringent limit at the short end of $\tau$ comes currently from observations of the light element abundances (shaded orange), although there is also a small region in which BBN constraints are outperformed by Ly$\alpha$ observations (shaded pink).

Interestingly, for a large range of neutralino masses subject to cosmological energy injection constraints---specifically, where the aforementioned scaling of the $m_{\chi_1^0}$ bound applies---the same constraints also rule out a massless axino ($m_{\tilde{a}} =0$).  
It is only in the degenerate ($m_{\chi_1^0} \simeq m_{\tilde{a}}$ and $\epsilon_{\rm em} \ll 0.5$) region where the excluded parameter space does not necessarily include $m_{\tilde{a}}=0$.
This conclusion is independent of the assumed value of $f_a'$ and stands in stark contrast to the gravitino superWIMP case, where the lower end of the exclusion region is always finite in $m_{\tilde{G}}$, whose precise value depends on our choice of $m_{\chi_1^0}$ (see Eq.~(\ref{eq:gravEx1}) and~(\ref{eq:gravEx2})).  Cosmological energy injection constraints therefore appear to have more drastic consequences for the axino than for the gravitino.  
 
At the long end of the decay lifetime ($\tau \sim 10^{12}-10^{23}$~s), the constraints again originate primarily 
from the {\it Planck} measurements of the CMB anisotropies (although we remind the reader here that the BBN limits have been cut off in this region only because of the lack of available calculations in the literature, while for the Ly$\alpha$ constraints our estimates of the free-streaming horizon may not suffice to model the observable effects of this parameter region; see Secs.~\ref{sec:BBN} and~\ref{sec:lya} for details).  The limits imposed by the CMB anisotropies at this end are most easily discernible in the right panels of Fig.~\ref{Fig:axinomass} showing the $(m_{\chi_1^0},\epsilon_{\rm em})$-plane (lower edge of red-shaded area), and 
translate generally into an upper limit on the axino mass.  Taking all observations into account, for the three values of $f_a'$ considered in this work and various fixed neutralino mass values, we find that cosmology constrains the axino mass $m_{\tilde{a}}$ to lie within the following ranges:  
\begin{itemize}
\item $f_{a}'=10^{16}$ GeV: 
\begin{equation}
\begin{aligned}
&m_{\chi_{1}^{0}}=100~{\rm GeV} &:&  \; 0\lesssim m_{\tilde{a}}/{\rm GeV} \lesssim 99.992, \\
&m_{\chi_{1}^{0}}=1~{\rm TeV} &:&  \; 0 \lesssim m_{\tilde{a}}/{\rm GeV} \lesssim 999.98 ,\\
&m_{\chi_{1}^{0}}=2.7~{\rm TeV} &:& \; 0 \lesssim m_{\tilde{a}}/{\rm GeV} \lesssim 2699.990,\\
&m_{\chi_{1}^{0}}= 4~{\rm TeV} &:&  \;  530 \lesssim m_{\tilde{a}}/{\rm GeV} \lesssim 3999.986,
\end{aligned}
\end{equation}

\item $f_{a}'=10^{12}$ GeV: 
\begin{equation}
\begin{aligned}
&m_{\chi_1^0}=100~{\rm GeV}  &:&  \; 98 \lesssim m_{\tilde{a}}/{\rm GeV} \lesssim 100-5 \times 10^{-5}, \\
&m_{\chi_{1}^{0}}=1~{\rm TeV} &:&  \; 998.9 \lesssim m_{\tilde{a}}/{\rm GeV} \lesssim 1000-8 \times 10^{-5}
\end{aligned}
\end{equation}

\item $f_{a}'=10^{8}$ GeV: 
\begin{equation}
\begin{aligned}
&m_{\chi_{1}^{0}}=10~{\rm GeV} &:& \;  0.00017 \lesssim m_{\tilde{a}}/{\rm GeV} \lesssim 0.000019.
\end{aligned}
\end{equation}
\end{itemize}
Thus, in terms of ruling out a large chunk of interesting particle masses in the GeV-to-TeV region conventionally of interest to collider searches, cosmological constraints have the strongest impact for large axion decay constants in the vicinity of $f_{a}'=10^{16}$~GeV, ruling out neutralino masses of up to $\sim 3$~TeV and all possible axino masses up to the degenerate limit.  The equivalent constraints in the same mass region are currently significantly weaker for smaller $f_a'$ values (i.e., $f_a'=10^{12}, 10^8$).  

Having said the above however, we also observe that only for $f_{a}'\lesssim 10^{8}$ GeV does the neutralino lifetime become short enough to be sensitive to LLP searches at the LHC (bottom left panel of Fig.~\ref{Fig:axinomass}, purple- and blue-shaded areas).   Specifically, for $f_a' = 10^8$~GeV, LLP searches at the LHC currently rule out neutralino and axino masses in the $\sim 50-500$ GeV region, which can be extended to $\sim 80-800$~GeV 
in the future by HL-LHC.  As in the gravitino case, reinterpretation of published ATLAS and CMS limits from the model-specific analyses of~\cite{ATLAS:2014kbb,CMS:2019ulr} (see Sec.~\ref{sec:gravitinoconstraints}) forces us to impose a cut-off on the LHC 
exclusion regions, this time at $m_{\tilde{a}} \sim 100$~GeV, which may or may not be indicative of the true sensitivity of the LHC searches. Nonetheless, 
as can be seen in Fig.~\ref{Fig:axinomass}, cosmological observations clearly have no sensitivity to the parameter combinations probed at colliders.  The complementarity between cosmological observations and collider searches is therefore self-evident.


\section{Conclusions}
\label{sec:conclusions}

We have revisited in this work cosmological constraints on Supersymmetric superWIMPs, and demonstrated that cosmology provides strong limits on two of the most well motivated candidates, the gravitino and the axino.

For the gravitino, the totality of cosmological constraints from energy injection and free-streaming considerations exclude neutralino and gravitino masses from a few eV to several TeV (Fig.~\ref{Fig:gravmass1}), effectively ruling out the bulk of the gravitino superWIMP parameter space. Measurements of the CMB anisotropies by the {\it Planck} CMB mission in particular extend the lifetime reach of cosmological observations to unprecedentedly long time scales (up to lifetimes of $\tau \simeq 10^{23}$~s, several orders of magnitude longer than the lifetime of the Universe thus far).  This enables us to close a large gap in the gravitino mass in the degenerate region ($m_{\tilde{G}} \simeq m_{\chi_1^0})$ previously allowed by observations of the light element abundances and CMB spectral distortions by COBE/FIRAS. Future CMB probes such has CMB-S4 and LiteBird will contribute to expanding this excluded region.  

In contrast, LLP searches at the LHC have a significantly smaller scope in terms of size of the gravitino superWIMP parameter space accessible to these searches, and the improvement expected from the upcoming HL-LHC is modest in comparison with the already vast parameter region covered by cosmological observations---up to masses of ${\cal O}(100)$~TeV, well beyond the kinematic reach of current and future collider experiments. Collider bounds are furthermore model-dependent and hinge on the neutralino production mechanism from cascade decay assumed in the analysis.  In contrast, cosmological constraints are conditioned only on the premise that a population of neutralino NLSPs is produced in some reasonable abundance in the early Universe and that this population decays electromagnetically.  Both assumptions can be easily satisfied independently of the model specifics. Notwithstanding their versatility, we emphasise that cosmological and collider bounds occupy completely different regions of the gravitino superWIMP parameter space, and certain parts of this parameter space remain as yet inaccessible to either. Thus, both types of probes remain as relevant as ever in the quest for BSM physics.

Similar conclusions hold for the axino superWIMP assuming $f_a'=10^{16}$~GeV, where again a large range of neutralino and axino masses up to the TeV scale are excluded, including $m_{\tilde{a}} =0$ (Fig.~\ref{Fig:axinomass}).  For lower values of the axino decay constant $f_a'$, the excluded region shifts to lower neutralino masses approximately as ${f_a'}^{2/3}$, such that for $f_a' \lesssim 10^{12}$~GeV the neutralino and axino masses excluded by cosmology typically fall below $\sim 10$~GeV.  We note however that collider constraints do not exist for $f_{a}' \gtrsim 10^{9}$~GeV, as the neutralino decay width is suppressed by ${f_a'}^{-2}$ and substantial decay occurring within the detector volume is simply too improbable for large $f_a'$ values.  Thus, while some parts of the axino superWIMP parameter space indeed lie within the reach of the LHC, our analysis here shows that within this class of models, the most model-independent and sweeping constraints in fact originate from cosmology. 

Lastly, while this work concerns SUSY superWIMPs, we note that the same cosmological constraints apply also to a large variety of well-motivated BSM scenarios that predict extremely weakly-coupled particles (either from symmetries or by the nature of the interactions) with limited signatures in conventional high-energy collider experiments.
These include models of cosmological relaxation~\cite{Graham:2015cka}, clockwork~\cite{Giudice:2016yja,Goudelis:2018xqi}, long-lived KK gravitons and radions~\cite{Feng:2003nr}, and continuum dark matter~\cite{Csaki:2021gfm}, etc.  The largely model-independent analysis presented in this work can be easily adapted to investigate how precision cosmological observations impact on such BSM scenarios.
We leave these studies for future works.

\section{Acknowledgements}
We thank Genevi\`{e}ve B\'{e}langer, Daniel R.~Green, Paul D.~Jackson, and Matteo Lucca for useful discussions.
Support for this work has been provided by the University of Adelaide and the Australian Research Council through the Centre of Excellence for Dark Matter Particle Physics (CE200100008).
Y$^3$W is supported in part by the Australian Government through the Australian Research Council’s Future Fellowship (FT180100031). 

\vspace{10mm}

\appendix
\section{The LSP momentum distribution}
\label{sec:appendixA}

Consider the decay process $\chi_1^0 \to {\rm LSP} + \gamma$, whose rate in the rest frame of $\chi_1^0$ (which coincides with the cosmic frame) is $\Gamma$.  The comoving number density of $\chi_1^0$  is 
\begin{equation}
n_{\chi_1^0}(t) = \frac{\Omega_{\rm DM} \rho_{\rm crit}}{m_{\chi_1^0}}    \exp\left(- \Gamma t\right),
\end{equation}
where $\rho_{\rm crit}$ is the present-day critical density, and we have assumed $\chi_1^0$ to be produced at an abundance that would match the observed reduced dark matter density $\Omega_{\rm DM}$ in the absence of decay.  Demanding that the rate of change of the comoving LSP number density $n_{\rm LSP}$ matches the negative rate of change of $n_{\chi_1^0}$, the evolution equation for $n_{\rm LSP}$ can be written as
\begin{equation}
\frac{{\rm d} n_{\rm LSP}}{{\rm d}t} = \Gamma n_{\chi_1^0} =\frac{\Omega_{\rm DM}  \rho_{\rm crit}}{m_{\chi_1^0}}  \Gamma  \exp\left(- \Gamma t\right),
\end{equation}
or equivalently in terms of the scale factor $R$, 
\begin{equation}
\label{eq:LSPcomovinglnR}
\frac{{\rm d} n_{\rm LSP}}{{\rm d}\ln R} =  \frac{\Omega_{\rm DM}  \rho_{\rm crit}}{m_{\chi_1^0}} 
  \left(\frac{R}{R_\Gamma} \right)^n  n \exp\left[- \left(\frac{R}{R_\Gamma} \right)^n\right], 
\end{equation}
where $n=2,\nicefrac{3}{2}$ applies to radiation domination and matter domination respectively, and $R_\Gamma$ is the scale factor at the time $t = 1/\Gamma$.

Observe that, at the moment of production, the LSP always carries a physical momentum $p_\star = \epsilon_{\rm em} m_{\chi_1^0}$, irrespective of the the time of of production.  Universal expansion, however, will cause the momentum to redshift, such that an LSP produced at a scale factor $R$ will be observed today uniquely with a physical momentum of $p=p_\star R$.  Thus, in the same way that we use $R$ as a proxy for time, we can also use the observed momentum $p$ as a proxy for $R$, and recast Eq.~(\ref{eq:LSPcomovinglnR}) as
\begin{equation}
\begin{aligned}
\label{eq:LSPcomovinglnp}
\frac{{\rm d} n_{\rm LSP}}{{\rm d}\ln p} = \frac{\Omega_{\rm DM}  \rho_{\rm crit}}{m_{\chi_1^0}} 
\left(\frac{p}{p_\star R_\Gamma} \right)^n  n \exp\left[- \left(\frac{p}{p_\star R_\Gamma} \right)^n\right].
\end{aligned}    
\end{equation}
Approximating the transition from radiation to matter domination to be instantaneous at $R_{\rm eq}$, 
LSP produced during radiation domination ($n=2$) can be observed today with a physical momentum up to $p_\star R_{\rm eq}$, while those arising from decay during matter domination ($n=\nicefrac{3}{2}$) will have momenta in the range $p_\star R_{\rm eq} < p \leq p_\star$. Integrating Eq.~(\ref{eq:LSPcomovinglnp}) under this approximation and assuming no pre-existing LSP population then yields Eq.~(\ref{eq:spectrum}).


\bibliographystyle{utphys.bst}
\bibliography{references}

\end{document}